\documentclass[12pt,reqno]{amsart}
\usepackage{graphicx}
\usepackage{amssymb,amscd,amsbsy}
\usepackage{amsthm}
\newcommand{\noi}{\noindent}

\newcommand{\C}{{\mathbb C}}
\newcommand{\RB}{{\mathbb R}}
\newcommand{\RP}{{\mathbb R}{\mathbb  P}}
\newcommand{\CP}{{\mathbb C}{\mathbb  P}}

\newcommand{\J}{{\mathcal J}}

\newcommand{\HH}{{\mathcal H}}
\newcommand{\MM}{{\mathcal M}}
\newcommand{\TT}{{\mathcal T}}
\newcommand{\LL}{{\mathcal L}}
\newcommand{\RR}{{\mathcal R}}

\def\On{\Omega_{N_0}}
\def\LM{L_{M}^{(2)}}
\def\LT{L_{\sigma}^{(2)}}
\def\inttt{\int\limits_{-2}^{2}}
\def\intinf{\int\limits_{-\infty}^{+\infty}}
\def\RT{\tilde{R}}
\def\ST{\tilde{S}}
\def\S{\mathcal S}

\def\Os{\Omega_0}

\def\U{\Upsilon}

\def\[{\left[}
\def\]{\right]}
\def\({\left(}
\def\){\right)}
\def\t{\theta}
\def\l{\lambda}

\def\s{\sigma}
\def\m{\mu}

\def\d{\delta}

\def\n{\nu}
\def\c{\chi}

\def\U{\Upsilon}
\def\b{\beta}
\def\z{\zeta}
\def\pb{\overline{p}}

\def\12{{1\over 2}}

\newcommand{\e}{{\boldsymbol e}}

\newcommand{\g}{\gamma}

\newcommand{\p}{\partial}
\newcommand{\eeq}{\end{equation}}
\newcommand{\beq}{\begin{equation}}
\newcommand{\bay}{\begin{eqnarray}}
\newcommand{\ey}{\end{eqnarray}}
\newcommand{\bey}{\begin{eqnarray*}}
\newcommand{\eey}{\end{eqnarray*}}

\newcommand{\sign}{\operatorname{sign}}

\newtheorem{thm}{\hspace{\parindent}Theorem}[section]

\newtheorem{lem}[thm]{\hspace{\parindent}Lemma}

\newtheorem{exa}[thm]{Example}

\theoremstyle{remark}
\newtheorem{rem}[thm]{Remark}
\newtheorem*{rem*}{Remark}

\begin{document}

\newcommand{\vse}{\vspace{.2in}}
\numberwithin{equation}{section}

\title{\bf Equations of  Camassa--Holm  type   and Jacobi ellipsoidal  coordinates}
\author{K.L.  Vaninsky}
\thanks{ The work is partially supported by NSF grant DMS-9971834.}

\begin{abstract}
We consider the integrable Camassa--Holm  equation on the line with positive initial data  rapidly decaying at infinity.   On such  phase space we     
construct a one parameter family  of integrable hierarchies  which preserves the mixed spectrum of  the associated   string spectral problem. 
This family  includes the CH hierarchy. 
We demonstrate that the constructed flows  can be interpreted as   Hamiltonian flows 
on the space of    Weyl functions of the associated  string spectral 
problem. The  corresponding Poisson 
bracket is the  Atiyah--Hitchin bracket. Using  an infinite dimensional version of the Jacobi ellipsoidal  coordinates we obtain  a one parameter 
family  of canonical  coordinates  linearizing the   flows. 
\end{abstract}
\maketitle
\setcounter{section}{0}
\setcounter{equation}{0}
\section{Introduction.}
\subsection{\bf Arnold's problem.}
Separation of variables is the simplest and the most powerful integration   method for equations  of motion  in  classical mechanics. Of course, there is no a general rule which allows one  to find such separating coordinates in the  general setting. ``Therefore we have to go in the opposite direction and knowing some remarkable substitution to find a problem where it can be  successfully applied", \cite{J}. One such remarkable substitution, known as Jacobi ellipsoidal  coordinates is widespread.  
Below  is a list (incomplete) of classical  problems where this substitution can be applied, \cite{AKN}. 
\begin{itemize}
\item Plane motion in the field of two attracting centers (Euler 1760). 
\item Kepler's problem in the homogeneous force field (Lagrange 1766). 
\item Geodesic motion  on  $n$-dimensional ellipsoid (Jacobi 1866).
\item Motion of uncoupled harmonic oscillators constrained to move on the $n$-dimensional sphere by the field of a quadratic potential (Neumann 1859).  
\end{itemize}

The following question\footnote{See also Problem 1985--16, \cite{A}.} was 
posed by V.I. Arnold:

\noi
{\bf Problem.  1981-29.}  \cite{A}. 
{\it Generalize Jacobi ellipsoidal coordinates to the infinite 
dimensional setting. Find equations of mathematical physics integrable by  this  method. } 

\noindent
The goal of the present paper is to demonstrate that the  Camassa--Holm equation is 
an example of a PDE integrable with this technique.  
The Camassa--Holm equation, \cite{CH}, is an approximation to the Euler equation 
describing  an ideal fluid
$$
{\p v\over \p t} + v{\p v\over \p x} +{\p  \over \p x}R\[v^2 +{1\over 2} \({\p v\over \p x}\)^2\]=0$$
in which $t\geq 0$ and $  -\infty < x < \infty$, $v=v(x,t)$ is velocity, and $R$ is inverse to $L=1-d^2/d x^2$ {\it i.e.}
$$
R[ f] (x)= {1\over 2} \int\limits_{-\infty}^{+\infty} e^{-|x-y|} f(y) dy.
$$
Introducing the function $m=L[v]$ one  writes the equation in the form
\footnote{We use   notation $D$ for the $x$-derivative and  $\bullet$ for the $t$--derivative. We use $\delta$ for the Frechet derivative.}
$$
m^{\bullet}+\(mD+Dm\)v=0. 
$$
The CH equation is a Hamiltonian system 
$m^{\bullet} +\{m,\HH\}=0
$
with  Hamiltonian
$$
\HH={1\over 2}\int_{-\infty}^{+\infty} mv\, dx=\text{energy}
$$
and the bracket
\beq\label{ppb}
\{A,B\}=\int_{-\infty}^{+\infty}{\delta A\over \delta m} \(mD+Dm\) {\delta B\over \delta m}\, dx.
\eeq

The integration of the CH equation is based on the spectral theory of an associated  Sturm--Liouwille operator. In connection with this spectral problem  Weyl, \cite{W},  in  1910  introduced what is  now called the Weyl function. 
The Weyl function depends on the  spectral parameter and defined such that  some  linear combination of fundamental solutions  and its derivatives vanishes at the end of an interval. 
The  novel point here is that the Weyl function carries a natural Poisson structure, the  Atiyah--Hitchin bracket. It corresponds to 
the Poisson structure \ref{ppb} 
on the phase space. The fact that Jacobi elliptic coordinates are  separating coordinates for \ref{ppb} is proved  by  
contour  integration.

It seems  that the CH equation on the line is the simplest among  all integrable PDE. The reason for this is  a degeneration of its spectral curve. 
Such degenerate reducible rational curves appear in the compactification of the space of smooth hyperelliptic curves. 
It is important that the constructions   of the present paper can be extended to other  basic integrable models, {\it i.e.}, the  KdV/KP and NLS equations, \cite{V2, V3}.  
 
The rest of the introduction describes  what is happening  for the CH equation. 
\subsection{\bf The  Camassa--Holm hierarchy.} 

We consider the CH equation with nonnegative ($m\geq 0$) initial data and such   decay at infinity that: 
$$
\int_{-\infty}^{+\infty} m(x) e^{|x|} dx < \infty.
$$
We denote this class of functions by $\MM$.
For such data  a solution of the initial value problem  exists for all times, see \cite{CM}. 
An initial profile moves from the left to the right and changes its shape under the flow.  We also need a subclass of functions $\MM_0 \subset \MM$ 
which vanish  far enough to the left. Evidently $\MM_0$  is invariant under the flow.

The CH Hamiltonian is just one of infinitely many conserved quantities of motion
\bey
\HH_1&=&\int v\,dx,\\
\HH_2&=&\frac{1}{2}\int v^2+(Dv)^2\,dx,\\
\HH_3&=&2\int v\[v^2+(Dv)^2\]\,dx,\\
\HH_4&=&{1\over 2} \int_{0}^{1}v^4+\int_{0}^{1}v^2(Dv)^2+2\int_{0}^{1}\[v^2+{(Dv)^2\over 2}\]G\[v^2+{(Dv)^2\over 2}\]\,dx,\; etc.
\eey
The first   integral 
$$
\HH_1=\int_{-\infty}^{+\infty} m\, dx=\text{momentum}
$$
produces the flow of translation $$m^{\bullet}+\{m,\HH_1\}=m^{\bullet}+Dm=0.$$
The second is the  CH Hamiltonian $\HH_2=\HH$.   
The  integrals $\HH_3,\; \HH_4,\; etc.,$ produce  higher flows of the CH hierarchy.  

An interesting feature of the Camassa--Holm equation is that the invariant manifold, or in other terms the spectral class  specified 
by these integrals, is not compact. As we will see this is reflected in the non-periodicity of the angle variables which occupy the entire real line.

\subsection{\bf The spectral problem. Action--angle variables}
One can associate to the CH equation an auxiliary string spectral problem, 
\footnote{We use prime $'$ to denote $\xi$-derivative.}
$$
f''(\xi)+ \l g(\xi)f(\xi)=0,\qquad\qquad -2\leq \xi \leq 2.
$$
The background information for  this spectral problem can be found in  \cite{GK, KK2, DM}. 
The variables $\xi$ and $x$ are related by 
$$
x\longrightarrow \xi =2\tanh {x\over 2}.
$$
Also the potential $g(\xi)$ is related to $m(x)$ by the formula $g(\xi)=m(x)\cosh^4{x\over 2}$. 
For initial data from $\MM$ the total mass of the associated string is finite $\int_{-2}^{+2} g(\xi)d\xi <\infty$. 
Evidently, for initial data from $\MM_0$ there is an interval of length $l$  where the potential vanishes:  
$g(\xi)=0,\quad\xi \in [-2,-2+l]$. 

The solutions  of the string spectral problem are continuous, but the left $f'_{-}(\xi)$ and right $f'_{+}(\xi)$ derivatives at some point $\xi$ may be different due to a concentrated mass at this point. 
Two solutions $\varphi(\xi,\l)$ and $\psi(\xi,\l)$ of the string spectral problem play an important role in the whole discussion. 
They are specified by  the initial data
\bey
\varphi(-2,\l)& =1\qquad\qquad\qquad \;   \psi(-2,\l)=0\\
\varphi'_{-}(-2,\l)&=0\qquad\qquad\qquad  \psi'_{-}(-2,\l)=1. 
\eey 

All potentials of the string spectral problem with  fixed  Dirichlet spectrum 
$$\l_n:\qquad \psi(2,\l_n)=0,\;\;\qquad\qquad\qquad n=1,2,\hdots;
$$ constitute  a  spectral class. The variables 
$$
I_n=-{1\over \l_n},\qquad\qquad\qquad \theta_n=\log \varphi(2,\l_n)
$$
are called ''action--angle" variables.  The angles  $\theta_n$ take values in $\RB$; they are not a cyclic variables.

It turns out that the  Dirichlet spectrum  is preserved under the flows of the CH hierarchy.
The flows  are linearized in terms of  angle variables. This follows from the canonical relations 
\bey
\{I_k,I_n\}&=0,\\
\{\theta_k,\theta_n\}&=0,\\
\{\theta_k,I_n\}&=\delta_k^n;\\
\eey
and the trace 
formulas of McKean, \cite{MC}:
\bey
\HH_1&=&\;\; \;\sum_{n}  \l_n^{-1}=-\sum_{n}  I_n,\\
\HH_2&=&{1\over 4}\sum_{n} \l_n^{-2}={1\over 4}\sum_n I_n^2,\qquad etc.
\eey
We  note that the CH flow was linearized in terms of the so--called coupling constants by Beals, Sattinger and Smigielski, \cite{BSS}.

\subsection{\bf Mixed boundary conditions} As we  noted  the CH flows do not change the Dirichlet spectrum. 
It is natural to consider the mixed boundary condition 
$$
a\varphi(2,\l)+b \psi(2,\l)=0,
$$
where $ a, b$ are real constants. The set of all boundary conditions, {\it i.e.}, the pairs $(a,b)$, constitutes  $\RP$, 
 real projective space. For example,  the Dirichlet boundary condition  corresponds to the point $(0,1)$. Our goal is to  associate 
to each boundary condition (point of $\RP$) the family of Hamiltonian flows  which preserve the spectrum corresponding to this boundary condition.    It turns out that 
the space  $\RP$ can be covered by two charts.

The first  chart  corresponds to the  boundary condition:
$$
\psi(2,\l)-C\varphi(2,\l)=0,
$$ 
in which $C\leq l$. It  can be constructed for all initial data from $\MM$. The family of roots $\l_n=\l_n(C),\; n=1,2,\hdots;$ produces generalized
''action--angle" variables  
$$
I_n(C)=-{1\over \l_n(C)};\qquad\qquad \theta_n(C)=\log\varphi(2,\l_n(C)).
$$
For each value of $C\leq l$ one has the family of  integrals. Here are the first two\footnote{The anti-derivative $D^{-1}$ is defined by the formula $D^{-1}f(x)=\frac{1}{2} \[
\int\limits_{-\infty}^{x}d\xi f(\xi)- \int\limits^{+\infty}_{x}d\xi f(\xi)\]$.} :
\bey
\HH_1(C)&=& -\frac{C}{4-C}\int_{-\infty}^{+\infty} m(x)(1+e^{-x}) dx + \frac{4}{4-C}\int_{-\infty}^{+\infty} m(x) dx; \\
\eey
\bey
4\HH_2(C)&=& \frac{1}{(4-C)^2}\(C \int_{-\infty}^{+\infty} m(x)(1+e^{-x}) dx - 4  \int_{-\infty}^{+\infty} m(x) dx   \)^2+\\
&+& \frac{2}{4-C}\times  \[\frac{C-4}{2} \(\;\;\int\limits_{-\infty}^{+\infty} d\xi m(\xi)\;\,\)^2-
2C \int\limits_{-\infty}^{+\infty} d\xi e^{-\xi}\,m(\xi) D^{-1}m(\xi)\right. +\\ 
 &\phantom{+}& \qquad\qquad\qquad\qquad\qquad\qquad\qquad\qquad+\left. (4-C)\int\limits_{-\infty}^{+\infty} d\xi m(\xi) v(\xi) \].
\eey
Evidently for  $C=0$ we have classical action-angle variables and the CH Hamiltonians.
The Hamiltonians produce the flows\footnote{The question of existence will be discussed at the end of  the introduction.}
$$
m^\bullet+\{m,\HH_k(C)\}=0,\qquad\qquad\qquad k=1,2,\hdots. 
$$
These flows preserve $I(C)$ and move $\theta(C)$ linearly . This follows from the trace formulas
\bey
\HH_1(C)&=&-\sum_{n}  I_n(C),\\
\HH_2(C)&=&{1\over 4}\sum_n I_n^2(C);\qquad \qquad\qquad 
\eey
and  the canonical relations for the variables $I(C)$ and $\theta(C)$.  

Another chart  corresponds to the boundary condition 
$$
\varphi(2,\l)+F\psi(2,\l)=0,
$$
in which $F > -1/l$. It is constructed for initial data from $\MM_0$.
For  $F=0$ the roots $\mu_k(0)$ form  the second spectrum of the string spectral problem 
$$
 \m_k:\;\qquad  \varphi(2,\mu_k)=0\qquad\qquad\qquad k=1,2,\hdots.
$$  
The roots $\mu_k(F),\;k=1,2,\hdots;$ produce $J(F)$ and $\tau(F)$,   the second family of canonical coordinates:
$$
J_k(F)=-{1\over \mu_k(F)},\qquad\qquad \tau_k(F)=\log \psi(2,\mu_k(F)).
$$
Again, for $F > -1/l$ one can write a family of  integrals. The first two are 
\bey
\TT_1(F)&=&\frac{1}{1+4F}\int_{-\infty}^{+\infty} m(x)(1+e^{-x}) dx + \frac{4F}{1+4F}\int_{-\infty}^{+\infty} m(x) dx;\\
4\TT_2(F)&=& \frac{1}{(1+4F)^2}\(\int_{-\infty}^{+\infty} m(x)(1+e^{-x}) dx +4F  \int_{-\infty}^{+\infty} m(x) dx   \)^2-\\
&-& \frac{2}{1+4F}\times  \[\frac{1+4F}{2} \(\;\;\int\limits_{-\infty}^{+\infty} dx\, m(x)\;\,\)^2-
2 \int\limits_{-\infty}^{+\infty} dx\, e^{-x}\,m(x) D^{-1}m(x)\right. -\\ 
 &\phantom{+}& \qquad\qquad\qquad\qquad\qquad\qquad\qquad\qquad-\left. (1+4F)\int\limits_{-\infty}^{+\infty} dx\, m(x) v(x) \].
\eey 
These Hamiltonians produce the flows 
$$
m^\bullet+\{m,\TT_k(F)\}=0,\qquad\qquad\qquad k=1,2,3,\hdots.
$$
The trace formulas have the form 
\bey
\TT_1(F)&=&-\sum_{n}  J_n(F),\\
\TT_2(F)&=&{1\over 4}\sum_n J_n^2(F),\qquad\qquad\qquad etc.
\eey

The formulas becomes  especially simple when   $F=0$. The first integral 
$$
\TT_1(0)=\int_{-\infty}^{+\infty} m(x) (1+e^{-x})\, dx,
$$
produces the  flow
$$
m^{\bullet}+\{m,\TT_1(0)\}=m^{\bullet}- 2me^{-x} +(1+e^{-x})Dm=0,
$$
which is linearized in the variables $J(0)$ and $\tau(0)$. The flows corresponding to  higher Hamiltonians are nonlocal.

\subsection{\bf Ellipsoidal coordinates. The Atiyah--Hitchin bracket} The key observation in the construction of the  generalized 
action--angle variables $I$ and $\theta$ and $J$ and $\tau$ is that the spectra $\l_n(C)$ and $\mu_k(F)$ can be interpreted as  Jacobi ellipsoidal coordinates. 
This  allows one to avoid any use of the  hierarchy  of differential equations  as well as the string spectral problem.  
As a result the dynamics  can be reformulated purely  in the language of meromorphic functions on $\CP$. 
First, let us  explain the    construction of $I(C)$. 

For the string $S_1$, {\it i.e.}, for the string with free left end and fixed right end,  we define so-called the Weyl function 
\beq\label{owf}
\Omega_0(\l)={\psi(2,\l)\over \varphi(2,\l)} = l+\sum_{k}{\sigma_k\over \mu_k-\l},
\eeq
where $l\geq 0$ is an interval free of mass,  $\sigma_k >0$ and $\sum \sigma_k/ \mu_k < \infty$. The roots of the equation $\Omega_0(\l)=C$, where $C \leq l$ are the points $\l_n(C)$. This is a classical way to define ellipsoidal coordinates. The construction  is depicted on Figure 1. 
 We have the sequence of points $\l_k=\l_k(C),\quad
k=1,2,\hdots$ which depend on the constant $C$. They interlace the poles of $\Omega_0(\l)$:
$$
\m_1<\l_1(C)<\mu_2<\l_2(C)<\hdots. 
$$
The action variables are defined by the formula
$$
I_n(C)=-\frac{1}{\l_n(C)},\qquad\qquad\qquad n=1,2,\hdots.
$$
Classical action variables correspond to the case   $C=0$.

\begin{figure}[h]
\includegraphics[width=0.60\textwidth]{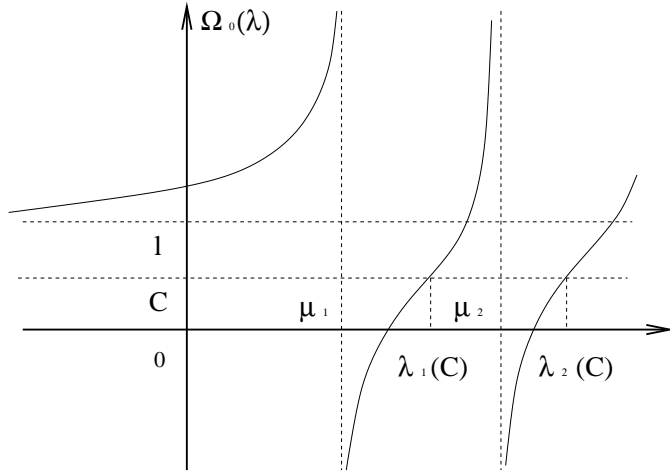}
\caption{The graph of function $\Omega_0$.}
\end{figure}

Now we can give a similar description of the variables  of $J(F)$. We assume that $m(x) \in \MM_0$ and  therefore $l >0$. For the string $S_0$ with fixed left and right ends we have the Weyl function 
\beq\label{ewf}
E_0(\l)=-{1\over \Omega_0(\l)}=-{\varphi(2,\l)\over \psi(2,\l)}=-{1\over l} +
\sum_{k}{\rho_k\over \l_k-\l},
\eeq
where $\rho_k >0$ and $\sum\rho_k/ \l_k <\infty$.  The points
$\mu_k(F)$ appear as  roots of the equation $E_0(\l)=F,\; F> -1/l$. This is shown on  Figure 2.

\begin{figure}[h]
\includegraphics[width=0.60\textwidth]{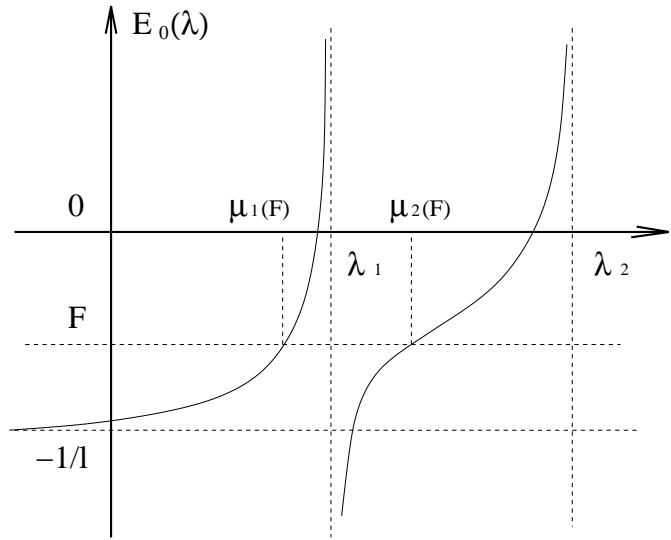}
\caption{The graph of function $E_0$.}
\end{figure}

\noi We have a sequence of points $\mu_k=\mu_k(F)$ interlacing the poles of $E_0(\l)$:
$$
\mu_1(F)<\l_1<\mu_2(F)<\hdots.
$$
The action variables $J(F)$ are defined by the formula
$$
J_n(F)=-\frac{1}{\mu_n(F)},\qquad\qquad\qquad n=1,2,\hdots.
$$
The second spectrum    corresponds to the case $F=0$. 

The next step is to relate the functions $\Os$ and $E_0$ to Poisson geometry. 
First we change the spectral parameter by the rule $\l\rightarrow \l'=-1/\l$. Note that this transformation maps the  spectrum into 
the corresponding action variables. 
The Poisson bracket \ref{ppb} for the function $\Omega_0$ corresponding two different values of the spectral parameter is given by the formula
$$
\{\Os(\l'),\Os(\m')\}={\(\Os(\l')-\Os(\m')\)^2\over \l'-\m'}. 
$$
This is  standard  Atiyah--Hitchin bracket in the form found by Faybusovich and Gehtman, \cite{FG}. The bracket (or more precisely the symplectic form) was originally introduced  for  rational functions in terms of their singularities in  \cite{AH}. In \cite{FG} coordinate free form of the bracket   for  rational functions was found. This coordinate free form is identically  the same  for an  infinite--dimensional case which we consider here. 
The bracket is invariant under linear--fractional transformations. Namely, 
for any function $\hat{\Omega}$ defined as 
$$
\hat{\Omega}={a\Os+b\over c\Os + d},
$$
where $a,b,c$ and $d$ are real constants, 
the Poisson bracket is given by the same formula, see \cite{V2}. In particular, the bracket 
for the function $E_{0}$ is given by the same formula.  The   
canonical  relations for the generalized 
action--angle variables  $I(C)$ and $\theta(C)$ and also $J(F)$ and $\tau(F)$  are  proved by contour integration.

\subsection{\bf The spectral curve and Baker--Akhiezer function}
It is instructive to \linebreak present an  algebraic--geometrical approach  to the inverse problems with pure discrete spectrum 
developed in \cite{KV,V1}. 
Such problems include finite Jacobi matrices \cite{KV,V1}, the quantum mechanical oscillator \cite{MCT} and the string spectral problem. 
The  algebraic--geometrical viewpoint will provide an additional geometrical insight into the nature of the flows constructed from the Hamiltonians $\HH(C)$ and $\TT(F)$.
  
The Riemann surface $\Gamma$ (Figure 3) associated   with the 
spectral problem consists of two   components $\Gamma_-$ and $\Gamma_+$, two copies of $\CP$.  A point on the curve is  denoted by $Q=(\l,\pm)$, where $\l\in \CP$ and the sign $\pm$ refers to the  
component.  An infinities of $\Gamma_{\pm}$ are  denoted by $P_{\pm}$ correspondingly. The components are glued together  at the points of the 
Dirichlet spectrum $(\l_k,\pm)$.

\begin{figure}[ht]
\includegraphics[width=0.80\textwidth]{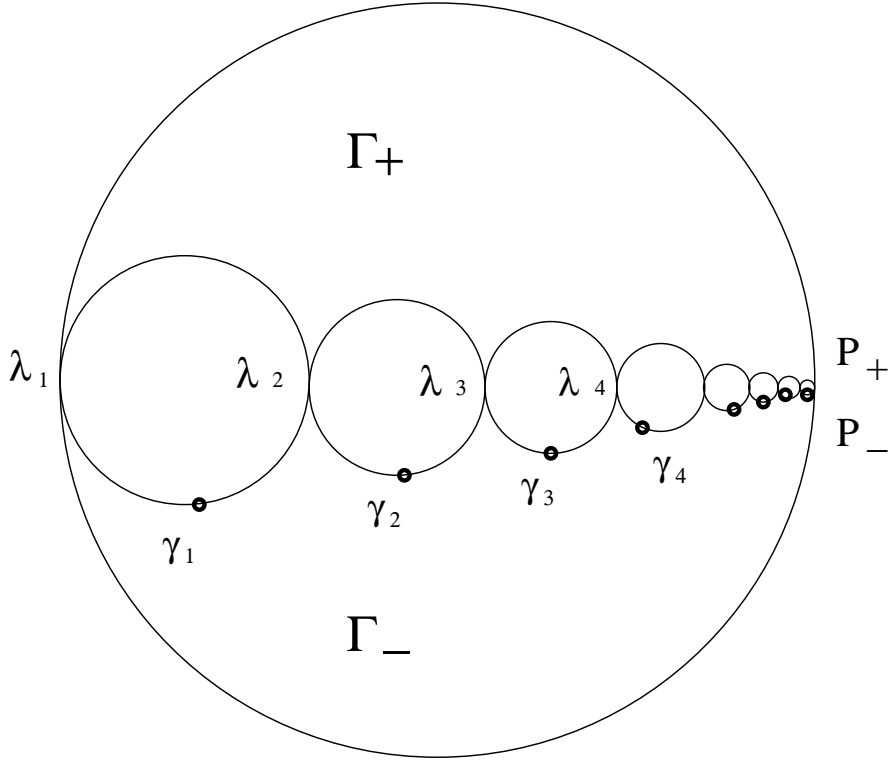}
\caption{The  Riemann surface.}
\end{figure}

The Baker--Akhiezer function $e(\xi,Q)$ is defined on $\Gamma$ and depends on the variable $\xi, -2\leq \xi \leq +2$ as a parameter. 
It is defined by the formula
$$
e(\xi,Q)= \left\{
\begin{array}
  {r@{\quad:\quad}l}
 \psi(\xi,\l) &       \qquad\qquad  \mbox{if}\;\;    Q\in \Gamma _+,\; \l=\l(Q);\\
{1\over E_0(\l)} \varphi(\xi,\l)+ \psi(\xi,\l), &  \qquad\qquad  \mbox{if}\;\;   Q\in \Gamma _-,\; \l=\l(Q);
\end{array} \right.  
$$ 
where
$$
E_0(\l)=-\frac{\varphi(2,\l)}{\psi(2,\l)}.
$$
The Baker--Akhiezer function has essential singularities  at the infinities $P_{\pm}$  and  simple poles at the points of the divisor $\gamma_k=(\mu_k,-),\; k=1,2,\hdots$. 
It is  holomorphic everywhere else. 
The Baker--Akhiezer function    satisfies the gluing condition 
$$
e(\xi,(\l_k,+))=e(\xi,(\l_k,-)),\qquad\qquad\qquad k=1,2,\hdots. 
$$ 
When $Q\in\Gamma _-$ we have 
$$
E_0(\l)={e'_-(-2,Q)\over e\,(-2,Q)},\qquad\qquad\qquad \l=\l(Q).
$$  
This type of formula expressing the Weyl function through the Baker--Akhiezer function is a starting  point for consideration 
in the case of hyperelliptic curves, \cite{V2, V3}.

The flows of the  Camassa--Holm hierarchy  constructed from the Hamiltonians $\HH(0)$ preserve the moduli $\l_k,\; k=1,2,\hdots,$ but move 
points of the divisor 
$\gamma_k,\; k=1,2,\hdots$.  The flows constructed from the Hamiltonians $\TT(0)$ preserve the divisor of the poles  and move the moduli. 
All other flows  constructed from the  Hamiltonians $\HH(C)$ and $\TT(F)$  move both the moduli and the divisor.

\subsection{\bf The spectral class.} Before going into the infinite--dimensional problem of the description of the spectral  class of a string   we present its finite--dimensional counterpart. Consider  a finite Jacobi matrix 
$$
L_0=\left[\begin{array}{ccccc}
v_0      & c_0       & 0         &\cdots        & 0\\
c_0      & v_1       & c_1       &\cdots        & 0\\
\cdot        &  \cdot    & \cdot     &\cdot         & \cdot\\
0    &  \cdots         & c_{N-3}         & v_{N-2}     & c_{N-2}\\
0       & \cdots    & 0         & c_{N-2}      & v_{N-1}
\end{array}\right],\qquad\qquad\qquad c_k>0.
$$
All matrices with the same spectrum  $\l_0<\ldots < \l_{N-1}$ as $L_0$
constitute a spectral class of $L_0$. There are no restrictions on the spectrum in this finite--dimensional case. 
The fact that the spectral class   is  diffeomorphic to $R^{N-1}$ 
was already noted by Moser, \cite{MO}. Tomei, \cite{To} showed that it 
can be compactified and  becomes a convex polyhedron. 
The symplectic interpretation 
of this result as a version of  the Atiyah--Guillemin--Sternberg
convexity theorem was given by Bloch, Flaschka  and Ratiu \cite{BF}.  
In \cite{V1} we  proved that points of the divisor can serve as  natural  coordinates 
on the spectral class  and described their range.  In this paper we give  a  similar description of the  string  spectral class in  
terms of the divisor.  

First we note that the two spectra 
\beq\label{ds}
\l_1<\l_2<\l_3<\hdots;
\eeq
and
\beq\label{ns}
\m_1<\m_2<\m_3<\hdots;
\eeq
appear as poles of the functions $E_0$  and $\Omega_0$   respectively. Their range is described by  two theorems of Krein.
The first theorem gives a description of all singular Riemann surfaces which might appear for initial data from $\MM$.

\begin{thm}\label{krlia} \cite{K}
 For a given increasing sequence  \ref{ds} to be the spectrum of a string $S_0$ it is necessary and sufficient that there exists a finite 
limit
$$
\lim_{n\rightarrow \infty} \frac{n}{\sqrt{\l_n}}
$$
and
$$
\sum_{n=1}^{+\infty} \frac{1}{\l_n^2 |D'(\l_n)|} <\infty,\qquad\qquad\qquad D(\l)=\prod_{n=1}^{+\infty} \(1-\frac{\l}{\l_n}\).
$$ 
\end{thm}

\noi
The next Theorem gives a description of all possible divisors. 
\begin{thm}\label{krmu} \cite{K}
For a given increasing sequence  \ref{ns} to be the spectrum of a string $S_1$ it is necessary and sufficient that there exists a finite 
limit
$$
\lim_{n\rightarrow \infty} \frac{n}{\sqrt{\m_n}}
$$
and
$$
\sum_{n=1}^{+\infty} \frac{1}{\m_n^{3/2} |D'_1(\m_n)|} <\infty,\qquad\qquad\qquad D_1(\l)=\prod_{n=1}^{+\infty} \(1-\frac{\l}{\m_n}\).
$$
\end{thm}
\noi
Now we can present a description   of the spectral class which is also due to Krein.  

\begin{thm}\label{tsc} \cite{KK2}
For two increasing interlacing sequences of $\l$'s and $\m$'s 
$$
0<\m_1<\l_1<\m_2<\hdots
$$
to be the  two spectra of a  string  with finite mass and length it is necessary and sufficient that
$$
-\sum_{k=1}^{\infty} \frac{1}{\m_k^2D_1'(\m_k)D(\m_k)}<\infty.
$$ 
\end{thm} 
\noi

Now  we have a complete description of  the spectral curve and divisor. We already introduced the formal flows corresponding to  
the  Hamiltonians $\HH(C)$ and $\TT(F)$.  We can say that the  flows   exist if the corresponding sequences 
 \ref{ds} and \ref{ns} satisfy the conditions of Theorem \ref{tsc}. To write an explicit form of the equations describing the 
motion of  the moduli and the divisor is a routine exercise. 

Finally, we  note that  the problem of the description of the spectral class for the quantum mechanical oscillator, \cite{MCT}, remains open.

\subsection{\bf The range of ellipsoidal  coordinates.} Now we want 
\begin{itemize}
\item given $\m$'s, to describe the range of variables $\l(C)$ when  $C <l$; 
\item given $\l$'s, to describe the range of variables $\mu(F)$ when  $F>-{1/l}$. 
\end{itemize}
\noi
An answer to these  questions  is given by two recent results. 

The first is the theorem of Kostuchenko--Stepanov, \cite{KS},  a direct generalization of an ancient (1857) formula of Boole, \cite{B}. In the context of the functions $E_0$ and $\Omega_0$ their result takes the following form. 

\noi
{\bf Theorem.}  
If $\sum_{k} \rho_k < \infty$ in  formula \ref{ewf} or $\sum_{k} \sigma_k <\infty$ in  formula \ref{owf}   then the   corresponding 
sequence of $\m_k(F)$ or $\l_k(C)$  satisfies 
$$
\sum_{k}\[ \l_k-\mu_k(F)\] <\infty,
$$ 
or
$$
\sum_{k} \[\mu_{k+1}-\l_k(C)\] < \infty. 
$$
The conditions of  the theorem are not only necessary but also sufficient for 
the sequence of  $\m$'s and $\l$'s   to arise as  coordinates for some  summable sequence of $\rho$'s or $\sigma$'s. 

 Unfortunately the 
condition of summability for the sequence of $\l$'s and $\m$'s is too restrictive for our purposes.  
It never holds for  initial data from the  phase space $\MM$. 
We need the second result which due to Nazarov and  Yuditskii, \cite{NY}.  

\noi
{\bf Theorem.}
If  $\sum\rho_k/ \l_k <\infty$ or $\sum \sigma_k/ \mu_k < \infty$, then 
$$
\sum_{k} {\l_k\over \mu_k(F)}-1 <\infty,
$$
or
$$
\sum_{k} {\mu_{k+1}\over \l_k(C)}-1 < \infty.
$$
The conditions of the  theorem are not only necessary but also sufficient for 
the sequence of  $\l$'s and $\m$'s  to arise as  coordinates for some  summable sequence of $\rho/\l$'s or $\sigma/\m$'s.

\subsection{\bf Organization of the paper}
In  Section 2 we introduce an auxiliary  Sturm--Liouville eigenvalue problem. Section 3  transforms  this spectral problem into the Krein string by  the classical Liouville transformation. 
We write the Poisson bracket  and the flows of the CH hierarchy in terms
of the potential of the string  spectral problem. Section 4 presents standard facts about the spectral theory of the string. 
Jacobi  ellipsoidal coordinates are discussed in Section 5. The Poisson bracket for the Weyl function is computed in Section 6. 
Action--angle coordinates are introduced in Section 7. There also CH flows are linearized. Section 8 contains examples of the computation of the Weyl 
function for one and two  peakon solutions. Section 9 introduces generalized action--angle coordinates which  include the  classical ones 
as a particular case.  Section 10 contains another family of canonical coordinates. Both families cover all possible spectra 
which arise from various mixed boundary conditions.  Section 11 presents the derivation of the  trace formulas. 

The author would like to thank Henry McKean for many constructive remarks. Many thanks to Mike Frazier for reading the 
manuscript and suggesting a lot of improvements. 

\section{ The spectral problem}
We consider the  Camassa--Holm  equation
\bey
{\p v\over \p t} + v{\p v\over \p x} +{\p p \over \p x}=0 \qquad \qquad
p(x)={1\over 2} \int\limits_{-\infty}^{+\infty} e^{-|x-y|}\[v^2 +{1\over 2} 
\({\p v \over \p x}\)^2\]\, dy,   
\eey
where $t\geq 0$ and $  -\infty < x < \infty$. The integral operator $R$ defined by 
$$
R[f](x)={1\over 2} \int\limits_{-\infty}^{+\infty} e^{-|x-y|} f(y)\,dy
$$  
is an inverse of  $L=1-D^2$. Introducing $m=L[v]$ 
we  specify the phase space  $\MM$ as all smooth positive functions with  such  decay at infinity  that 
$$
\int\limits_{-\infty}^{\infty} m(x)e^{|x|} \,dx < \infty.
$$
We also need $\MM_0 \in \MM$ the subspace of functions which vanish  far enough  to the left.  
$\MM$ and $\MM_0$ are invariant under the CH flow, \cite{CM}.

The first four Hamiltonians, integrals of motion, are listed in 1.2. The first Hamiltonian $\HH_1=\int\limits_{-\infty}^{+\infty} v\, dx$=momentum  produces the flow of translation: 
\beq\label{htf}
0=m^{\bullet}+ \{m,\HH_1 \}=m^{\bullet}+ Dm,       
\eeq
where the bracket is
\beq\label{pb}
\{A,B\}=\int\limits_{-\infty}^{+\infty} {\d A\over \d m} (mD +Dm) 
{\d B\over \d m} \, dx.  
\eeq
The CH Hamiltonian  $\HH_2={1\over 2} \int v^2 +{Dv}^2 \, dx$=energy  is the  second in the infinite series of commuting integrals of motion. The equation can be written as 
\beq\label{chh}
0=m^{\bullet} + \{m, \HH_2\} = m^{\bullet}+  \( mD +Dm\) v=
({\p\over \p t}+2(Dv) +vD) (1-D^2)v.  
\eeq
All other integrals produce higher flows of the hierarchy.

The CH equation  is a compatibility condition between 
\beq\label{slsp}
D^2 f-{1\over 4}\,  f + \l m f=0  
\eeq
and 
$$
f^{\bullet} =-\(v+{1\over 2\l}\) Df +{1\over 2} (Dv) f;
$$
{\it i.e.}, $(D^2 f)^{\bullet} =D^2(f^{\bullet})$ is the same as \ref{chh}. The 
isospectrality of  \ref{slsp} is a key to the integrability of the CH dynamics.

\section{The Liouville correspondence}
The standard Liouville's  transformation 
$$
x \rightarrow \xi(x)=2 \tanh x/2,\qquad\qquad
f(x) \rightarrow f(\xi)={f(x)\over \cosh x/2}
$$ 
converts  \ref{slsp}   into the string  spectral problem     
\beq\label{ssp}
f''+ \l g f =0,    
\eeq
with $g(\xi)=m(x)\cosh^4 x/2$ and $-2 \leq \xi \leq +2$. 
The transformation changes the length element  by the rule 
$$
dx\longrightarrow d\xi = dx \, \J(\xi), \qquad\qquad \text{with} 
\qquad \J(\xi)= 1-{\xi^2\over 4}.
$$
Under the assumptions made, the phase space of the string is regular,  {\it i.e.}, its mass is finite: 
$$
\int\limits_{-2}^{+2} g(\xi) \, d\xi=\int\limits_{-\infty}^{+\infty} 
m(x) \cosh^2 x/2 \, dx < \infty.
$$
The  transformation  reduces the problem \ref{slsp} with two singular ends to the regular string  
problem on the finite interval. 
It is necessary for  introduction of the ellipsoidal coordinates but first   we consider what is happening on the phase space. 

The functions $m(x)$ and $v(x)$ are related by $m=L[v]$ and $v=R[m]$. 
Introducing $k(\xi)=v(x) \cosh^4 x/2$, we have the  relations  
$$
g(\xi)=\LL [k](\xi) = k(1-4 {\J'}^ 2-2 \J \J'') - 5 k'\J \J'-k'' \J^2, 
$$
and 
$$
k(\xi)=\RR [g](\xi) = {1\over 2}\int\limits_{-2}^{+2} \RR(\xi,\eta) g(\eta) \, 
d\eta 
$$
with 
$$
\RR(\xi,\eta)=\left\{ \begin{array}{c}
\({1+\eta/2\over 1+ \xi/2}\)^2  {1\over \J(\xi) },  \qquad\qquad \qquad 
\text{for $\eta \leq  \xi$};\\
                 \(1-\eta/2\over 1-\xi/2\)^2   {1\over \J(\xi)},  \qquad \qquad\qquad\text{for $\eta \geq  \xi$}.
                 \end{array}
                 \right.  
$$
The functions
$k(\xi)$ and $g(\xi)$ are   new coordinates 
on the phase space. 
Best of all, the relation between the functions $m,v$ and $g,k$ can be expressed by the diagram: 
$$
m(x): \;  m =L[v]  \qquad\qquad   \longleftrightarrow  \qquad\qquad   v(x):\; v =R[m]
$$
\vskip 0.1in
$$
g(\xi)=m(x)\cosh^4 {x\over 2}\quad\updownarrow \qquad\qquad\qquad\qquad  \qquad  \qquad\qquad\qquad\updownarrow  \quad  k(\xi)=v(x)\cosh^4 {x\over 2}
$$
\vskip 0.1in
$$
g(\xi): \;  g=\LL[k]    \qquad\qquad \quad    \longleftrightarrow  \qquad\qquad     k(\xi): \; k = \RR[g]
$$
\vskip 0.3in
\noi

Now we want to express the first two flows \ref{htf} and \ref{chh} of the  CH hierarchy in terms of the new coordinate, the 
density function $g$. The conserved quantities  are 
\bey
\HH_1&=&\int\limits_{-2}^{+2} g\,  \J   d\xi, \,   \\
\HH_2&=&{1\over 2} \int\limits_{-2}^{+2} g\,  \RR [g] \J^3 d \xi,\quad etc. 
\eey
The bracket \ref{pb}  takes the form\footnote{ $\partial$ 
stands for the derivative in $\xi$ variable.}
\beq\label{npb}
\{A,B\}=\int\limits_{-2}^{+2} {\d A\over \d g} \( \J g \partial \J^{-1} + 
\J^{-1} \partial g \J\) {\d B\over \d g}\, d\xi.
\eeq 
Computing the  variation $\d\HH_1=\int \d g\, \J \, d\xi$  we 
write the translation flow \ref{htf}  as 
$$
0=g^{\bullet}+ \{g,\HH_1\}= g^{\bullet}+ \J^{-1}\partial \(g \J^2\).
$$
The self-adjointness of $\RR$ acting   on  the functions $f$ and $g$ of the  
$\xi$--variable, 
$$
\int\limits_{-2}^{+2}    \RR[g]\,  f \J^3\, d\xi =
\int\limits_{-2}^{+2}  g \, \RR[f]  \J^3 \, d\xi
$$
implies the formula for the  variation of the basic  Hamiltonian 
\bey
\d \HH_2 &= {1\over 2} \int \d g \RR[g]\, \J^3 d\xi 
+ {1\over 2} \int g \RR[\d g] \, \J^3  d\xi
   = \quad \int \d g \, \RR[g] \,  \J^3 d\xi.  
\eey
Therefore, for the CH flow \ref{chh},  we have 
$$
0= g^{\bullet} + \{g, \HH_2\}= g^{\bullet}+ (g\J\partial \J^{-1}+\J^{-1}\partial g\J) k \J^3.
$$
Evidently the Liouville's transformation complicates the formulas for the flow. 

\section{ The spectral theory of the string} 
In this section we introduce the so--called Weyl function, the main tool in  construction of ellipsoidal coordinates. We explain its relation to the spectral theory of the string. The direct and inverse spectral theory of the string with nonnegative 
mass  was constructed by M.G. Krein 
in the 1950's and presented in \cite{KK2}, see also \cite{GK,DM}.  To formulate the results we need some facts of function theory. These can be found 
in  \cite{KK1}. 

The function of a complex variable $F(\l)$ belongs to the class $(R)$ and is called  an 
$R$--function if it is holomorphic on $\C-\RB$ with $F(\overline{\l})=\overline{F(\l)}$ and $\Im F(\l)$ has the same sign as $\Im \l$.
Any such function can be represented in the form 
\beq\label{irb}
F(\l)=\alpha+\beta \l +\intinf\({1\over \zeta-\l}-\frac{\zeta}{1+ \zeta^2}\)d\s(\zeta)\qquad\qquad (\Im \l \neq 0),
\eeq
where $\beta \geq 0$ and $\alpha$ is real constant. The positive measure $d\s$ is such that
$$
\intinf \frac{d\s(\zeta)}{1+\zeta^2} <\infty.
$$
 Denote by $(\RT)$ the class $(R)$ adjoined with  the function identically equal to infinity. Evidently, if $F(\l)\in (\RT)$ then 
 $-1/F(\l) \in (\RT)$.

Let  $(R_1)$ be a subclass   of functions which admit an 
absolutely convergent integral representation 
\beq\label{ir}
F(\l)=\alpha+\intinf{d\s(\zeta)\over \zeta-\l}\qquad\qquad (\Im \l \neq 0),
\eeq
where $\alpha$ is a real constant and $d\s$ is a positive measure.

We introduce two solutions $\varphi(\xi,\l)$ and $\psi(\xi,\l)$ of 
the eigenvalue problem \ref{ssp} with the boundary conditions
\bey
\varphi(-2,\l)& =1\qquad\qquad\qquad \;   \psi(-2,\l)=0\\
\varphi'_{-}(-2,\l)&=0\qquad\qquad\qquad  \psi'_{-}(-2,\l)=1. 
\eey
The functions $\varphi(2,\l)$  and  $\psi(2,\l)$  can be written in the form
\beq\label{prod}
\varphi(2,\l)=\prod_{k=1}^{\infty}\(1-{\l\over \m_k}\),\qquad\qquad \psi(2,\l)=4\prod_{k=1}^{\infty}\(1-{\l\over \l_k}\). 
\eeq
The roots interlace each other
$$
0<  \m_1 <\l_1 < \m_2 < \l_2 <\hdots.
$$

Given some function $N(\l)$ which belongs to the function class $(\RT)$. This function describes the boundary condition at the right end of the string. 
For example, if the right end of the string is fixed then $N(\l)=0$ and {\it vice-versa}.
The function $\Omega_N$  was introduced by 
H. Weyl, \cite{W},  in his study of the Sturm--Liouville problem. It is defined such  that the corresponding  
Weyl solution 
$
\chi=-\psi + \Omega_N \varphi
$
satisfies  boundary condition 
$$
\chi'_+(2)N(\l) + \chi(2)=0 \qquad\text{ for all } \;  \l. 
$$
The Weyl function is the main ingredient of the spectral theorem.

To formulate the spectral 
theorem  we consider the measure $dM(\xi)=g(\xi)d\xi$ where $g(\xi)\geq 0$ is the  function from \ref{ssp} and  $\LM=\LM[-2,2]$ the set of $M$--measurable 
complex--valued functions $f(\xi)$ such that  
$$
||f||_{M}^2=\inttt |f(\xi)|^2 dM(\xi)< \infty. 
$$
Evidently, $\LM$ is a Hilbert space.  For any nondecreasing function 
$\s(\l)$ given on the interval $(-\infty,+\infty)$,  we denote by 
$\LT=\LT(-\infty,+\infty)$ the set of $\s$--measurable functions $F(\l)$ such that
$$
||F||_{\s}^2=\intinf |F(\l)|^2 d\s(\l) < \infty.
$$

The string with a fixed left end is denoted by $\S_0$ and  the string 
with a free left end is denoted by $\S_1$. 
Consider the string  $\S_1$ first. The nondecreasing function $ \s$ is 
called a spectral function of a string $\S_1$ if  the mapping $U:\; f\rightarrow F$, where $f\in \LM[-2,+2]$,  and
$$
F(\l)=\inttt f(\xi) \varphi(\xi,\l)dM(\xi),
$$
isometrically carries the space $\LM[-2,+2]$ into the Hilbert space $\LT$. 
Accordingly the Parseval equation 
$$
\intinf |F(\l)|^2d\s(\l)= \inttt|f(\xi)|^2 \, dM(\xi)
$$
holds. The spectral function $\s$ of the  string is said to be orthogonal 
if the mapping  $U:\quad f\rightarrow F$ maps $\LM$ onto the entire space $\LT$.

We say  that the string has  heavy endpoints if  $\xi=-2/+2$ are
points of increase/decrease  of the function $M(\xi)$.

\begin{thm}
 {\bf The Spectral Theorem.} Suppose  the string $\S_1$ 
has heavy endpoints and does not carry a concentrated mass 
at the right  endpoint. Then the formula 
$$
\Omega_N(\l)={\psi'_+(2,\l) N(\l) + \psi(2,\l)\over 
\varphi'_+(2,\l)N(\l) +\varphi(2,\l)}\qquad\qquad\qquad (\Im \l 
\neq 0)
$$
defines an $R_1$--function for any choice of $N(\l)\in (\RT)$. Any such function admits the integral representation \ref{ir} with $\alpha=0$:
$$
\Omega_N(\l)=\intinf\frac{d\s_N(\zeta)}{\zeta -\l}.
$$
These formulas establish a one-to-one correspondence $ N\leftrightarrow \s$  
between the class $(\RT)$ of functions $N(\l)$ and the set  of all spectral functions $\s(\l)$ of the string $\S_1$.  
The spectral function $\sigma_N(\l)$ will be orthogonal if and only if the function $N(\l)$ corresponding to it degenerates into a real constant, 
possibly  infinity.  
\end{thm} 

To state some facts about strings  we need two additional subclasses of $(R)$. 

We say that the function $F(\l)\in (R)$ belongs to the class $(S)$ if it is holomorphic on $(-\infty,\,0]$ and $\geq 0$ there. 
 All such functions have an absolutely convergent integral representation 
\beq\label{irs}
F(\l)=\alpha+\int^{+\infty}_{-0}{d\s(\zeta)\over \zeta-\l},
\eeq
where $\alpha\geq 0$ and $d\s$ is a positive measure. Evidently $(S)\subseteq (R_1)$. Denote by $(\ST)$ the class $(S)$ adjoined with the function identically equal to infinity.

Similar,  the function $F(\l)\in (R)$ belongs to the class $(S^{-1})$ if it is holomorphic on $(-\infty,\,0]$ and $\leq 0$ there.  
All such functions have an  integral representation 
\beq\label{irss}
F(\l)=\alpha+\beta \l +\int^{+\infty}_{+0}\({1\over \zeta-\l}-{1\over \zeta}\) d\rho(\zeta),
\eeq
where $\alpha \leq 0,\;\beta \geq 0$ and $d\rho$ is a positive  and such that 
$$
\int_{+0}^{+\infty} \frac{d\rho(\zeta)}{\zeta +\zeta^2}        < \infty .
$$
 Denote by $(\ST^{-1})$ the class $(S^{-1})$ adjoined with the function identically equal to infinity.
It can be proved that  $F(\l)\in (\ST)$ if and only if  
 $-1/F(\l) \in (\ST^{-1})$.

\begin{rem}\label{weylsol} Let the right end of the string is fixed ($N(\l)=0$). 
For such a string $\S_1$ the Weyl function $\Omega_0 \in (\ST)$ and 
according to the general formula \ref{irs} 
$$
\Omega_0(\l)={\psi(2,\l)\over \varphi(2,\l)}=\sum_{k=1}^{\infty}{\s_k\over \mu_k-\l}.
$$
\end{rem}

\begin{rem}\label{weylfree}
If the left end of the string $\S_1$ contains an interval   
free of mass then the Weyl function takes the form \ref{irs} with $\alpha=l$ where  the constant $l$ is the  length  of 
the interval.  Therefore, for the string $\S_1$ with fixed right end, we have 
\beq\label{wfl}
\Omega_0(\l)=l+ \sum_{k=1}^{\infty}{\s_k\over \mu_k-\l}. 
\eeq
From  formula \ref{prod} we have 
$$
\Omega_0(0)={ \psi(2,0)\over\varphi(2,0)}= 4.
$$ 
Combining this with  formula \ref{wfl} we obtain  
\beq\label{wfz}
l+ \sum_{k=1}^{\infty}{\s_k\over \mu_k}=4. 
\eeq
\end{rem}

\begin{rem}\label{weylinv}
 Given  $\m$'s which are spectra of the string $\S_1$, what $l$ and what sequence of $\sigma$'s can appear in  formula \ref{wfl}? 
For any  sequence of $\sigma$ such that 
$$
\sum_{k=1}^{\infty}{\s_k\over \mu_k}< \infty, 
$$
one has a function $\Omega_0 \in (\ST)$ given by  formula \ref{wfl}. If $l$ is such that $\Omega_0(0)=4$ (what is equivalent to \ref{wfz}), then 
to such $\Omega_0$ corresponds a unique regular string $\S_1$ with heavy right end and the interval  at the left of length $l$ free of mass.  
\end{rem}

\begin{rem}\label{weylfu}
 For the string $\S_0$ with  fixed left end, the set of all 
Weyl  functions is given by the formula 
$$
E_{N}=-{1\over \Omega_N}.
$$
If the right end is also fixed then $E_0(\l)\in (\ST^{-1})$ and   according to general theory  
\beq\label{gggg}
E_0(\l)=-{1\over \Omega_0(\l)}=-{\varphi(2,\l)\over \psi(2,\l)}= -{1\over 4}+\sum_{k=1}^{\infty} \(\frac{1}{\l_k-\l}-\frac{1}{\l_k}\)\rho_k.  
\eeq
The poles of $E_0(\l)$ are zeros of $\psi(2,\l)$  and as we will see they do not 
move under the CH flow. 
The sum
$$
\sum_{k=1}^{\infty} \frac{\rho_k}{\l_k} <\infty 
$$ 
if and only if  $l>0$. For such a string $\S_0$,  the formula \ref{gggg} becomes 
\beq\label{gg}
E_0(\l)=-{1\over l}+\sum_{k=1}^{+\infty} \frac{\rho_k}{\l_k-\l}.
\eeq
Evidently 
\beq\label{ggh}
-{1\over l}+\sum_{k=1}^{+\infty} \frac{\rho_k}{\l_k}=-\frac{1}{4}
\eeq
\end{rem}

\begin{rem}\label{weylspec}

 Given the sequence of $\l$'s which is the spectrum of the string $\S_0$, what $l$ and sequence of $\rho$ can appear in  formula \ref{gg}?
For any  sequence of $\rho$ such that 
$$
\sum_{k=1}^{\infty}{\rho_k\over \mu_k}< \infty
$$
one has a function $E_0 $ defined by  formula \ref{gg}. If $l$ is such that $E_0(0)=-\frac{1}{4}$ (what is equivalent to \ref{ggh}), then 
$E_0(-x) < E_0(0) <0$ for any $x>0$. Therefore, 
$E_0 \in  (\ST^{-1})$. The function 
$$
\Omega_0=-\frac{1}{E_0}.
$$
belongs to $ \ST$. 
The function $\Omega_0$ has  the form \ref{wfl} and satisfies the condition $\Omega_0(0)=4$. According to Remark 4 there exists a unique corresponding regular string $\S_1$ with heavy right end and the interval  at the left free of mass.  
\end{rem}
\section{ Jacobi ellipsoidal coordinates}

This section contains information about Jacobi  coordinates in finite and infinite dimensions, see 
\cite{J} lecture 26. The  results  are used to describe the range of the  canonical variables constructed 
in the subsequent sections. The present discussion (and notations) are  independent from all other sections.

First we explain the construction of Jacobi   coordinates in finite dimensions. Consider 
the function 
$$
\U(\l)=\sum_{k=1}^{N}{\n_k\over \gamma_k-\l},\qquad\qquad\qquad\nu_k >0.
$$
Assume that the poles $\gamma_1 <\gamma_2 <\hdots <\gamma_N$ are ordered and fixed. The residues $\nu_1,\hdots,\nu_N$ are 
considered as variables  and they fill up $R^N_+$. Pick $C\neq 0$ and consider points where $\U(\l)=C$. 
We  have $N$ real roots $\c_k=\c_k(C),\;k=1,\hdots,N$. These are Jacobi  coordinates.   If $C >0$  then,
\beq\label{iso}
\c_1< \gamma_1 <\c_2 <\hdots <\c_N < \gamma_N;  
\eeq
and if $C <0$, then 
\beq\label{ist}
\g_1 < \c_1< \g_2 <\hdots  < \g_N <\c_N.  
\eeq
The converse is also true. Pick some $C$ and consider the points $\c_k,\;k=1,\hdots,N$ which satisfy 
inequalities \ref{iso} or \ref{ist} correspondingly. There exist a unique sequence of 
$\n_1,\hdots, \n_N$ from $R^N_+$ corresponding to the sequence of $\c$'s. 

The following formula\footnote{A. Volberg pointed this to me.} is due to Boole, \cite{B}: 
\beq\label{bi}
C\sum_{k=1}^{N}\g_k-\c_k=s_0, \qquad\qquad\quad  \qquad C\neq 0, 
\eeq
where 
$$
s_n=\sum_{k} \n_k \g_k^n,\qquad\qquad\qquad n=0,1,\hdots.
$$
In fact Boole's formula is the first in the  infinite sequence of identities. The second that we found is  
$$
C^2\sum_{k=1}^{N}\g_k^2-\c_k^2= 2s_1C-s_0^2,\qquad\qquad\qquad\qquad\qquad C\neq 0.
$$
One can continue  and obtain the formulas 
for 
$$
\sum_{k=1}^{N}\g_k^p-\c_k^p
$$
in terms of the moments $s_0,\hdots,s_{p-1}$ for any integer $p$. 
We present two proofs of Boole's formula. The first is classical and can be found in  \cite{L}.

The identity
$$
C-\U(\l)=C{\prod (\l-\c_k)\over \prod (\l-\g_k)}
$$
implies 
$$
C\prod_{k}(\l-\g_k)-\sum_{k}\n_k\prod_{p\neq k} (\l-\g_p)=C\prod_{k}(\l-\c_k).
$$
The coefficients of  $\l^N$  of both sides  match. Matching 
coefficients of  $\l^{N-1}$ we obtain \ref{bi}. 

The second proof is a little more complicated, but it can be easily generalized. 
Note that near infinity,
$$
\U(\l)=-{s_0\over \l}-{s_1\over \l^2}-{s_2\over \l^3}-\hdots.
$$
Consider the contour integral 
$$
{1\over 2\pi i}\int \l{\U'(\l)\over \U(\l)-C}\;d\l
$$
 over a circle  of sufficiently large radius. 
The poles and zeros of the function $\U-C$ produce a contribution to the integral equal to 
$$
\sum \g_k-\c_k.
$$
From another side, near infinity 
$$
{\U'(\l)\over \U(\l)-C}=-\l^{-2}s_0/C-\l^{-3}(2s_1/C-s_0^2/C^2)-\hdots.
$$
Thus the integral is equal to 
$$s_0/C
$$ which is  Boole's formula.

The second and all higher identities   can be proved by contour integration with $\l$ replaced by $\l^p,\; p\geq 1$.

Consider now the infinite--dimensional  case. Let
$$
\U(\l)=\sum_{k=1}^{\infty} {\n_k\over \g_k-\l},
$$
where $\g_k\rightarrow + \infty$ as $k\rightarrow \infty$ and $s_0=\sum \n_k < \infty$. For any 
$C$ we have a sequence of $\c$'s which interlace the sequence of $\l$'s similarly  to 
\ref{iso} or \ref{ist}. Pick some $C\neq 0$. Can any sequence of $\c$'s interlacing  $\l$'s  arise from some sequence of $\n$'s similarly to the finite--dimensional case?  
In fact,  there is an asymptotic condition on $\c's$  which   follows from the next result. 
\begin{thm}\cite{KS}
The $\c$'s  are  coordinates, i.e. they arise from some $C \neq   0$ and some 
summable sequence of $\n_s$ if and only if 
$$
\sum_{k=1}^{+\infty} |\g_k-\c_k| <\infty.
$$
If this holds then
$$
C\sum_{k=1}^{+\infty} \(\g_k-\c_k\)= s_0.
$$
\end{thm}
\noindent
This result  can be obtained  from the rational case in the limit   when  
the  number of poles tends  to infinity.

Unfortunately the condition $s_0 <\infty$ does not  hold in  most cases we consider. 
For the string with light left end ({\it i.e.} without concentrated mass) we have only   
$$
s_{-1}=\sum_{k=1}^{+\infty}{\n_k\over \g_k} <\infty.
$$
\begin{thm}  \cite{NY} Pick some $C >0$. 
The $\c$'s  are ellipsoidal coordinates, i.e. they arise from  some 
 sequence of $\n_s$  with  $s_{-1} <\infty$   if and only if 
$$
\sum_{k=1}^{+\infty}\( {\g_k\over \c_k} - 1\)<\infty. 
$$
If $C <0$ , then the above condition should be replaced by
$$
\sum_{k=1}^{+\infty} \({\c_k\over \g_k} - 1\)<\infty. 
$$
\end{thm}

{\it Proof.} Without loss of generality we assume that $C=\U(0) >0$. Then, 
$$
\U(\l)-\U(0)=\exp F(\l)
$$ 
where $F\in (R)$. Using \ref{irb} we have 
$$
\U(\l)-\U(0)=A\exp\({\int f_0(\zeta){1+\zeta\l\over \zeta-\l} {d\zeta\over 1+\zeta^2}}\),
$$
where 
$$
f_0(\l)=\left\{ \begin{array}{ll}
1 &  \qquad\qquad\qquad \mbox{for  $\qquad\g_k <\l <\c_{k+1},\;\quad k=0,1,\hdots;$}\\
0 &   \qquad\qquad\qquad \mbox{for $\qquad\c_k <\l <\g_k,\;\qquad k=1,2,\hdots.$}
         \end{array}  \right.
$$ 
We put formally $\g_0=-\infty$. 
Writing   $f_0=1-f$, we have
$$
\U(\l)-\U(0)=-A\exp\({-\int_{0}^{+\infty} f(\zeta){1+\zeta\l\over \zeta-\l} {d\zeta\over 1+\zeta^2}}\).
$$ 
Now pass to the limit with $\l=-x \longrightarrow -\infty$ along the real axis. Then the  limit 
$$
\lim_{x\rightarrow +\infty}\int_{0}^{\infty} f(\zeta){\zeta x -1\over \zeta +x} 
{d\zeta\over 1+\zeta^2}= \int_{0}^{+\infty} f(\zeta){\zeta\over \zeta^2 +1} d\zeta < \infty 
$$
exists. 
This implies 
$$
\sum_{k=2}^{\infty} \log {\g_k\over \c_k} < \infty, 
$$
which is equivalent to the condition of the theorem. Moreover,
$$
\U(\l)-\U(0)=-\U(0)\exp\({-\int_{0}^{+\infty} {f(\zeta)d\zeta\over \zeta-\l}}\). 
$$

Conversely, pick $C >0$ and  the sequence of $\c$'s  which interlace the sequence of $\g$'s:
$$
0=\c_1<\g_1<\c_2<\hdots, 
$$
and satisfy the condition of the theorem. 
Define the function $\U(\l)$ by the identity 
$$
\U(\l)=C -C\exp\({-\int_{0}^{+\infty} {f(\zeta)d\zeta\over \zeta-\l}}\), 
$$
where
$$
f(\zeta)=\sum_{k=1}^{\infty} 1_{[\c_k,\g_k]}(\zeta).
$$
Evidently, $\U(\l)$ maps the upper half--plane into itself. It also vanishes at infinity and has poles at $\g_k$. Thus, applying \ref{irb} again 
$$
\U(\l)=\alpha + \beta \l  +\sum_{k=1}^{\infty} {1+\g_k\l\over \g_k-\l} {\n_k\over 1+\g_k^2}. 
$$
Now pass to the limit with $\l \longrightarrow -\infty$ along the real axis. The  existence of the 
limit implies that $\beta=0$ and 
$$
\sum_{k=1}^{\infty} {\g_k\n_k\over1+\g_k^2} <+ \infty.
$$
Therefore,  since $\U(\l)$ vanishes at infinity, 
\bey
\U(\l)&
=&\alpha+
\sum_{k=1}^{\infty}\( {1\over \g_k-\l}
-{\g_k\over 1+\g^2_k}\)\n_k\\&
=&\(\alpha-\sum_{k=1}^{\infty}{\g_k\n_k\over 
 1+\g^2_k} \)+\sum_{k=1}^{\infty}{\n_k\over \g_k-\l}\\
&=&\sum_{k=1}^{\infty}{\n_k\over \g_k-\l}.\qquad\qquad\qquad\qquad \blacksquare
\eey

When $C>0$ we have $\c_k < \g_k$ for all $k$, and the condition  
$$
\sum_{k=1}^{+\infty}\( {\g_k\over \c_k} - 1\)<\infty 
$$
is equivalent
$$
\prod_{k=1}^{+\infty} {\g_k\over \c_k} <\infty. 
$$
This product can be expressed in the spirit of Bool's formula as
$$
\prod_{k=1}^{+\infty} {\g_k\over \c_k}=\frac{C}{C-s_{-1}}.
$$

To prove the formula we consider an algebraic situation with $N < \infty$. We assume $C > \U(0)=s_{-1}$ and this implies that $\c_1>0$. The function 
$$
\log (\U(\l)-C)
$$
is single valued on the plane cut along the segments $[\c_k,\g_k],\; k=1,\hdots, N$.
Consider the contour integral 
$$
{1\over 2\pi i}\int \frac{1}{\l}\log (\U(\l)-C)\;d\l
$$
over a circle  of sufficiently large radius. Evaluating this integral when the radius tends to  infinity, we have 
$$
\log (-C).
$$
Evaluating it in the finite part of the plane we have to account a contribution from the origin and  small circles surrounding the cuts. 
At the origin
$$
{1\over 2\pi i}\int \frac{1}{\l}\log (\U(\l)-C)\;d\l=\log (s_{-1}-C),
$$  
and over a  circle surrounding $[\c_k,\g_k]$
$$
{1\over 2\pi i}\int \frac{1}{\l}\log (\U(\l)-C)\;d\l= -{1\over 2\pi i}\int \log  \l \frac{\U'(\l)}{\U(\l)-C}\;d\l=\log \g_k- \log \c_k. 
$$
Thus 
$$
\log (-C)=\log (s_{-1}-C) +\sum_{k=1}^{N} \(\log \g_k- \log \c_k\). 
$$
The result follows by exponentiating this equation. The formula extends  to all other values of $C$ and to functions $\U$ with an arbitrary number of poles. We do not dwell on this. 

\section{ Poisson bracket for the  Weyl function} 

The goal of the  present section is to compute the  Poisson bracket \ref{npb} 
for the Weyl function. 
\begin{thm}\label{ddd} Let $N(\zeta)=N_0$ be a real constant, possibly infinity. 
Then,
\beq\label{ppo}
\{\On(\l),\On(\m)\}={\l\m\over \l-\m}\(\On(\l)-\On(\m)\)^2.  
\eeq
\end{thm}

This fact is central for all further discussion. 
The proof follows the steps developed in \cite{V2} for the Dirac operator. We start with two auxiliary lemmas.

\begin{lem}\label{grad} The gradient of $\On(\l)$ is 
$$
{\p \On(\l)\over \p g(\xi)}=\l \chi^2(\xi),
$$
where $\chi=-\psi+\On \varphi$.
\end{lem}
{\it Proof. } The proof is now standard, see \cite{V2}.

Let $f=f(\xi,\l)$ and $s=s(\xi,\m)$ 
be  arbitrary solutions of the string spectral problem
$$
f''+\l gf=0,\qquad\qquad\qquad s''+\m gs=0.
$$
\begin{lem}\label{iden} Let  $\J=\J(\xi)$ be an arbitrary function. Then
$$
s^2g \J\p \, {f^2\over \J}- f^2g \J\p \, {s^2\over \J}= {1\over \m-\l}\p\[f's-f s'\]^2.
$$
\end{lem}

 This formula can be verified directly.

{\it Proof of Theorem \ref{ddd}.} Using  formula \ref{npb} for the Poisson 
bracket,  Lemma \ref{grad} and \ref{iden} and integration by parts,  we compute  
\bey
\{\On(\l),\On(\m)\}&=&\int_{-2}^{+2}{\d\On(\l)\over \d g(\xi)}
\( \J g \partial \J^{-1} + \J^{-1} \partial g \J\){\d\On(\m)\over \d g(\xi)}\, d\xi\\
&=&\l\m\int_{-2}^{+2}\chi^2(\l)
\( \J g \partial \J^{-1} + \J^{-1} \partial g \J\)\chi^2(\m)\,d\xi\\
&=&-\l\m\int_{-2}^{+2}\chi^2(\m) g\J\partial \,{\chi^2(\l)\over \J} - 
\chi^2(\l) g\J\partial \,{\chi^2(\m)\over \J}\, d\xi\\
&=&-{\l\m\over \m-\l}\[\chi'(\l)\chi(\m)-\chi(\l)\chi'(\mu)\]^2|^{+2}_{-2}.
\eey
To compute the value at the point $\xi=2$ we write
$$
\chi'(\l)\chi(\m)-\chi(\l)\chi'(\mu)=\[{\chi'(\l)\over \chi(\l)}-{\chi'(\m)\over \chi(\mu)}\]
\chi(\l)\chi(\mu).
$$
Note that $\chi' N_0 +\chi=0$ at $\xi=2$. Therefore, at this point 
$$
{\chi'(\l)\over \chi(\l)}={\chi'(\m)\over \chi(\m)}=N_0,
$$
and  the contribution from the upper limit vanishes.  

At the lower limit $\xi=-2$, using $\chi=-\psi+\On \varphi$ we have 
$$
\chi(\l)=\On(\l),\qquad\qquad \qquad \chi'(\l)=-1.
$$
Finally,
$$
=-{\l\m\over \m-\l}\[\chi'(\l)\chi(\m)-\chi(\l)\chi'(\mu)\]^2|_{-2}=
{\l\m\over \l-\m}\(\On(\l)-\On(\m)\)^2.
$$
The proof is finished.

 If one changes the spectral parameter by the rule 
$$
\l' = -{1\over \l},\qquad\qquad\qquad \m' = 
-{1\over \m},
$$ 
then the  formula of the  theorem becomes
\beq\label{ahb}
\{\On(\l'),\On(\m')\}={\(\On(\l')-\On(\m')\)^2\over \l'-\m'}.
\eeq
This is the  standard  Atiyah--Hitchin bracket, \cite{AH}, in the form found by Faybusovich and Gehtman, \cite{FG}. Note 
that the   function $\On(\l')$  maps the upper half--plane into itself. Due to the  invariance 
 of the AH bracket under linear--fractional transformations, \cite{V2}, the bracket for the 
function $E_{N_0}(\l')$ is given by the same formula \ref{ahb}.

\section{ Action--angle variables. Dynamics of the Weyl function}

The main result of this section is 

\begin{thm}\label{tcp} The variables 
$$
I_n=-{1\over \l_n},\qquad\qquad\qquad \theta_k=\log \varphi(2,\l_k)
$$
are ''action--angle" variables
\bey
\{I_k,I_n\}&=0,\\
\{\theta_k,\theta_n\}&=0,\\
\{\theta_k,I_n\}&=\delta_k^n.\\
\eey
\end{thm}

We start the proof with two  auxiliary lemmas. 
In section 4 we introduced  the representation \ref{gggg} for the function $E_0(\l)$:
$$
E_0(\l)= -{1\over 4}+\sum_{k=1}^{\infty} \(\frac{1}{\l_k-\l}-\frac{1}{\l_k}\)\rho_k.  
$$
Introducing
\beq\label{one}
\l_k'=-{1\over \l_k},\qquad\qquad \rho_k'={\rho_k\over \l_k^2},   
\eeq
we have the identity
\beq\label{two}
{\rho_k\over \l_k-\l}={\rho_k'\over \l_k'-\l'}+ {\rho_k\over \l_k}.  
\eeq
Therefore   for the function $E_0(\l')$  we obtain
\beq\label{three}
E_0(\l')= -{1\over 4} + \sum_{k=1}^{\infty} {\rho_k'\over \l_k'-\l'}.
\eeq

First, we compute the bracket between the  the parameters $\l_k'$ and $\rho_k'$ entering 
into  formula \ref{three}.
\begin{lem}\label{ocp} The bracket  \ref{ahb} in $\l'$ and $\rho'$ coordinates has 
the form
\bay
\{{\rho_k}',\rho_n'\}&=&{2 \rho_k' \rho_n'\over \l_n'-\l_k'}(1-\delta_k^n),  \label{fi} \\
\{\rho_k',\l_n'\}&=&\rho_k'\delta_k^n,  \label{si} \\
\{\l_k',\l_n'\}&=&0. \label{ti} 
\ey
\end{lem}
\noindent
The proof is identical to the proof of Theorem 2 in  \cite{V1}.

Now we can compute the bracket between the   parameters $\l_k$ and $\rho_k$ entering 
into  formula \ref{gggg}.
\begin{lem}\label{oocb} The bracket  \ref{ppo} in $\l$ and $\rho$ coordinates has 
the form
\bay
\{\rho_k,\rho_n\}&=&{2\l_k \l_n \rho_k \rho_n\over \l_k-\l_n},\label{fid}\\
\{\rho_k,\l_n\}&=&\l_k^2 \rho_k \delta_k^n, \label{sid}\\
\{\l_k,\l_n\}&=&0.\label{tid}
\ey
\end{lem}

The proof follows  from the results of the previous lemma.

{\it Proof of Theorem \ref{tcp}.} The first identity follows trivially from  \ref{tid}. Commutativity of angles 
is more complicated. Using \ref{prod}
\bey
\{\t_n,\t_m\}&=& \sum_{k,k'}\;\{\log\(1-{\l_n\over \m_k}\),\log\(1-{\l_m\over \m_{k'}}\)\}\\
&=&\sum_{k,k'} {1\over  \(1-{\l_n\over \m_k}\) \(1-{\l_m\over \m_{k'}}\)} \{{\l_n\over \m_k},
{\l_m\over \m_{k'}}\}\\
&=&\sum_{k < k'} {\l_n\l_m \over  \(1-{\l_n\over \m_k}\) \(1-{\l_m\over \m_{k'}}\) {\m_k}^2 
{\m_{k'}}^2} \(\{{ \m_k,\m_{k'}}\} + \{{ \m_{k'},\m_{k}}\}\)\\
&=&0.
\eey
The last sum vanishes due to skew symmetry of the bracket.

To prove the last relation we use the identity $\varphi(2,\l_k)=\psi'(2,\l_k)\rho_k$. We have 
\bey
\{\t_k,I_n\}&=&\{\log \psi'(\l_k)+\log\rho_k, -{1\over \l_n}\}\\
&=&{1\over \rho_k\l_n^2}\{\rho_k,\l_n\}={1\over \rho_k\l_n^2} \l_k^2\rho_k \d^n_k=\d^n_k.
\qquad\qquad\blacksquare
\eey

H. McKean, \cite{MC}  expressed the conserved quantities $\HH_1$ and $\HH_2$ in terms of the Dirichlet 
spectrum
\bey
\HH_1&=&\int\limits_{-\infty}^{+\infty} m\, dx= \int\limits_{-2}
^{+2} g \J d\xi= \sum_{n=1}^{+\infty}  \l_n^{-1},\\
\HH_2&=&{1\over 2}\int\limits_{-\infty}^{+\infty} m R[m]\, dx=
{1\over 2}\int\limits_{-2}^{+2} g \RR[g]\J^3 \, d\xi= 
{1\over 4}\sum_{n=1}^{+\infty} \l_n^{-2}.
\eey
These are obtained by matching coefficients of two  expansions near $\l=0$ 
for the solution $\psi$. One expansion is obtained from  formula \ref{prod}; the another form  a Neumann series\footnote{These and more general trace formulas 
will be obtained in Section 11.}.

The evolution of residues under the translation flow
$$
\rho_k^{\bullet}=\{\rho_k,\HH_1\}=\sum_{n}\{\rho_k, \l_n^{-1}\}= 
- \l_k^{-2}\{\rho_k,\l_k\}=-\rho_k,
$$
{\it i.e.}
$$
\rho_k(t)=\rho_k(0)e^{-t}.
$$
Therefore, under the translation flow 
$$
E_0(\l)=-{1\over 4}+ \sum_{k=1}^{\infty} \({1\over \l_k-\l}- \frac{1}{\l_k}\)\rho_k(0)e^{- t}.
$$

The evolution of residues under the CH flow obeys
$$
\rho_k^{\bullet}=\{\rho_k,\HH_2\}={1\over 4}\sum_{n}\{\rho_k, 
\l_n^{-2}\}= - {1\over 2} \l_k^{-3}\{\rho_k,\l_k\}=-\pb_k \rho_k,
$$
where $\pb_k={1\over 2\l_k}$. Therefore, 
$$
\rho_k(t)=\rho_k(0)e^{-\pb_k t},
$$
and
$$
E_0(\l)=-{1\over 4}+ \sum_{k=1}^{\infty} \({1\over \l_k-\l}- \frac{1}{\l_k}\)\rho_k(0)e^{-\pb_k t}.
$$

\section{ Examples. 1 and 2 peakon solutions}

A remarkable class of solutions of the CH equation  are  
peakon--antipeakon solutions of the form $v(x,t)=\sum_{n} 
p_n e^{-|x-q_n|}$. The parameters $q_n$ and $p_n$  satisfy the Hamiltonian 
flow 
$$q^{\bullet}_n=\p H/ \p p_n, \;\qquad\qquad  p^{\bullet}_n=-\p H/
\p q_n$$ 
with  Hamiltonian $H={1\over 2}\sum p_i p_j 
e^{-|q_i-q_j|}$ and the classical Poisson bracket.  

It is instructive to compute the  evolution of Weyl functions for the simplest one and two  peakons solutions.  
The equations of motion of two peakons were already integrated in \cite{CHH}, but the evolution of the Weyl function  computed here is an illustration  
of results of the previous section.  Since everything is algebraic we allow $p_n$ to be of both signs. 

\begin{exa}{\bf One peakon.} \end{exa} \noi The solution has the form $v(x,t)=p(t) e^{-|x-q(t)|}$, 
with $p(t)=p_0$ and $q(t)=q_0+ pt$. The  peakon ($p >0$) travels 
to the right while the anti--peakon ($p< 0$) travels to the left.

The solutions $\varphi$ and $\psi$ take the form
\bey
\varphi(2,\l)&=&-\l m_1l_1 + 1,  \\
\psi(2,\l)&=& -\l m_1 l_0 l_1 +4.
\eey
Using $m_1l_0 l_1=8p$ for the Weyl function of the string $S_0$ with 
fixed right end ($N\equiv 0$) we  obtain 
$$
E_0(\l)= - {\varphi(2,\l)\over \psi(2,\l)}=-{1\over l_0} + {
\rho(t)\over {1\over 2p} -\l},
$$
with
\beq\label{ef}
\rho(t)={1\over m_1l_0^2}= {1\over 8p} e^{-q}={e^{-q_0}\over 8p}e^{-pt}. 
\eeq

\begin{exa}{\bf Two peakons.} \end{exa} \noi
The solution is of the form 
$$v(x,t)=p_1(t)e^{-|x-q_1(t)|}+ 
p_2(t)e^{-|x-q_2(t)|}.
$$ 
The two--particle Hamiltonian is 
$$
H={1\over 2} p_1^2 + {1\over 2} p_2^2  + p_1 p_2 e^{-|q_1-q_2|} =\12 \pb_1^2 
+ \12\pb_2^2,
$$
where $\pb_n$ is the  asymptotic velocity of the $n$-th particle. The equations 
of motion 
\bey
q_1^{\bullet}&=&{\p H\over \p p_1}=p_1+p_2e^{-|q_1-q_2|},\qquad 
p_1^{\bullet}=-{\p H\over \p q_1}= \; p_1p_2 \; \sign\, 
(q_1-q_2) e^{-|q_1-q_2|},\\
q_2^{\bullet}&=&{\p H\over \p p_2}=p_2 + p_1 e^{-|q_1-q_2|},\qquad 
p_2^{\bullet}=-{\p H\over \p q_2}= -p_1p_2\; \sign\, (q_1-q_2) e^{-|q_1-q_2|};  
\eey 
are integrated by introducing the new variables
\bey
P&=p_1+p_2, \qquad\qquad Q=q_1+q_2,\\
p&=p_1-p_2, \qquad\qquad q=q_1-q_2.
\eey
This reduces reduces the equations of motion to 
\bey
P^{\bullet}&=&0, \qquad\qquad\qquad\qquad \qquad\qquad\qquad Q^{\bullet}=P(1+e^{-|q|}),  \\
p^{\bullet}&=&{1\over 2} (P^2-p^2)\, \sign\,  q e^{-|q|}, \qquad\qquad\quad
q^{\bullet}=p\, (1-e^{-|q|}). 
\eey
Now the Hamiltonian  is 
$$
H={1\over 4}\[P^2+p^2+(P^2-p^2)e^{-|q|}\]. 
$$

Consider the case\footnote{The case $\pb_1=\pb_2$ is excluded by the 
equations of motion.} $\pb_1 >  \pb_2 >0$. This is the pure peakon case.  
The law of conservation of energy takes the form 
$$
{4H-P^2-p^2\over P^2-p^2}=e^{-|q|},
$$
with  left hand side   less then $1$ for all times.  
This implies that $q=q_1-q_2 <0$ for all times. The solution is
\bey
p(t)& =&A{1-\b e^{At}\over 1  + \b e^{At}},\\
q(t)&=& \log {A^2\b  e^{At}\over (\pb_1+\pb_2\b e^{At})(\pb_2+\pb_1\b e^{At})},\\
Q(t)&= &Q(0)+Pt -\log{\pb_1+\pb_2\b e^{At}\over \pb_2+\pb_1\b e^{At}}+ \log
{\pb_1+\b\pb_2\over \pb_2+\beta\pb_1}; 
\eey 

with
$$
A=\pb_1-\pb_2,\qquad\qquad \b={A-p(0)\over A+p(0)}> 0.
$$

The solutions $\varphi$ and $\psi$ are 
\bay
\varphi(2,\l)&=&  \l^2 l_1 l_2 m_1m_2 -\l(m_1l_1+m_1l_2+m_2l_2)+ 1, \label{es} \\ 
\psi(2,\l)&= &\l^2 m_1 m_2 l_0 l_1 l_2 -\l(m_1l_0 l_2 + m_1l_0 l_1+m_2l_0 l_2+m_2l_1 l_2) +4.  \label{et} 
\ey
Note that   
$$
m_1m_2l_0 l_1 l_2 =8P^2-16H=16\pb_1 \pb_2
$$
and
$$
m_1l_0 l_2+m_1l_0l_1+m_2l_0l_2+m_2l_1l_2=8P=8(\pb_1+\pb_2).
$$
Therefore,  for the poles $(\l:\; \psi(\l)=0)$ we obtain
$$
\l_1={1\over 2\pb_1}\qquad\qquad \l_2={1\over 2\pb_2}.
$$
Thus:
$$
E_0(\l)=-{\varphi(2,\l)\over \psi(2,\l)}=-{1\over l_0}+ {\rho_1(t)\over 
{1\over 2\pb_1}-\l} +{\rho_2(t)\over {1\over 2\pb_2}-\l}
$$
with
\bay
\rho_1(t)&=&{\varphi(2,{1\over 2\pb_1})\over 8 (\pb_2-\pb_1)}=
{R\over 8A} e^{-\pb_1t}, \label{eff}\\
\rho_2(t)&=&{\varphi(2,{1\over 2\pb_2})\over 8(\pb_1-\pb_2)}= 
{R\over 8A}\b e^{-\pb_2t} \label{efi}
\ey
and 
\beq\label{esi}
R^2={e^{-Q(0)}\over \b} {\pb_2+\b \pb_1\over \pb_1+\b\pb_2}. 
\eeq

{\bf Peakon-antipeakon.}
Now to the  case $\pb_1 >0 >\pb_2$. The integration of the equations of motion is identical to the pure peakon case. 
The solution is 
\bey
p(t)&=&A {1+\z e^{At}\over 1-\z e^{At}},\\
q(t)&=&\log{A^2\z e^{At}\over (\pb_1\z e^{At}-\pb_2)(\pb_1-\pb_2\z e^{At})},\\
Q(t)&=&Q(0) +Pt -\log{\pb_1-\pb_2\z e^{At}\over \pb_1-\pb_2 \z} + 
\log{\pb_2-\pb_1 \z e^{At}\over \pb_2- \pb_1 \z}; 
\eey
where
$$
\z={p(0)-A\over p(0) +A}> 0.
$$
Thus,
$$
E_0=-{\varphi(2,\l)\over \psi(2,\l)}=-{1\over l_0} +{\rho_1(t)\over 
{1\over 2\pb_1} -\l}+{\rho_2(t)\over {1\over 2\pb_2}-\l},
$$
with
$$
\rho_1= {D\over 8 A}e^{-\pb_1t},\qquad\qquad \rho_2(t)=
-       {D\over 8A}\z e^{-\pb_2t},
$$
and
$$
D^2={e^{-Q(0)}\over \z}{ \pb_1\z -\pb_2\over \pb_1-\pb_2\z}.
$$
The formulas for residues are obtained in the same way as in the 
pure peakon case. At the moment of the collision the mass of the first 
particle tends to infinity and $s_1=\rho_1(t)+\rho_2(t)=0$.

\section{ Generalized action--angle variables}

The action--angle variables introduced in  section 7 are a particular case of the following construction. According to  formula \ref{wfl}, the function $\Omega_0(\l)$ has the form
$$
\Omega_0(\l)={\psi(2,\l)\over \varphi(2,\l)}= l+ \sum_{k=1}^{\infty}{\s_k\over \mu_k-\l},\qquad\qquad\qquad l\geq 0. 
$$
All possible sequences of $\mu$'s are described in Theorem \ref{krmu}.
 
Pick some  constant $C$ which does not exceed $l$, and find the points where $\Omega_0(\l)=C$ 
(see Figure 1).  
 We have the sequence of points $\l_k=\l_k(C),\quad
k=1,2,\hdots$ which depend on the constant $C$. They interlace the poles of $\Omega_0(\l)$:
\beq\label{inl}
0< \m_1<\l_1<\mu_2<\l_2<\hdots. 
\eeq
As it was explained in section 5, these are Jacobi ellipsoidal coordinates. Section 7 describes the case when $C=0$. Given $\mu$'s,  what is the range of all possible sequences of $\l(C)$'s? 

The  Nazarov--Yuditskii theorem implies 
$$
\sum_{k=1}^{\infty} \(\frac{\l_k(C)}{\mu_k}-1\) <\infty.
$$
Conversely, let  the sequence of $\l$'s  satisfy \ref{inl} and the above condition.  Then again by the Nazarov--Yuditskii theorem there exists a 
function 
$$
\U(\l)=\sum_{k=1}^{\infty} \frac{\sigma_k^0}{\mu_k-\l},\qquad\qquad\qquad\sum_{k=1}^{\infty} \frac{\sigma_k^0}{\mu_k} <\infty;
$$
such that $\U(\l_k)=-1,\quad k=1,2,\hdots$. Note that $\U(0)>0$  depends only on the given sequence of $\mu$'s and $\l$'s. 
Pick $l,\; 0\leq l< 4$ and define 
\beq\label{ips}
\Omega_0(\l)=l+ \frac{(4-l)\U(\l)}{\U(0)}.
\eeq
By construction
$$
\Omega_0(\l_k)= l- \frac{(4-l)}{\U(0)}=C,\qquad\qquad\qquad k=1,2,\hdots
$$
the range of $C$ being the interval $-4/\U(0)\leq C<4$. The function $\Omega(\l)$ has the form \ref{wfl} and satisfies  condition \ref{wfz}. Therefore, according to remark \ref{weylinv}, for any choice of $l$, or equivalently $C$, 
there exists a string $\S_1$ with the function $\Omega_0$ given  by  formula \ref{ips}.

\begin{rem}
 For $C=0$,  there  exists a string $\S_1$ with $l>0$ and corresponding Dirichlet spectrum coinciding with the $\l$'s. 
\end{rem}

Now we want to 
construct variables canonically conjugate to $\l(C)$.

\begin{thm}\label{ttcp} The variables 
$$
I_n(C)=-{1\over \l_n(C)},\qquad\qquad\qquad \theta_k(C)=\log \varphi(2,\l_k(C))
$$
are canonically conjugate variables
\bey
\{I_k(C),I_n(C)\}&=0,\\
\{\theta_k(C),\theta_n(C)\}&=0,\\
\{\theta_k(C),I_n(C)\}&=\delta_k^n.\\
\eey
\end{thm}
To prove the theorem we introduce the function $E_0^c(\l)$:
$$
E_0^c=-{1\over \Omega_0-C}.
$$ 
 It has the representation 
$$
E_0^c(\l)=-{\varphi(\l)\over \psi(\l)-C\varphi(\l)}=-{\varphi(\l)\over \psi^c(\l)}.
$$
The points $\l_k(C)$ are the roots of the equation $\psi^c(\l)=0$.
The function $\psi^c(\l)$ can be written as 
$$
\psi^c(\l)=(4-C)\prod_{k=1}^{\infty} \(1-{\l\over \l_k(C)}\).
$$
 Evidently $4-C >0$, since $C\leq l <4$. 
The function  $E_0^c(\l)$   maps the upper half--plane into itself and can be written as:
\beq\label{nf}
E_0^c(\l)= -{1\over l-C}+\sum_{k=1}{\rho_k(C)\over \l_k(C)-\l}. 
\eeq
One can change the variable $\l\longrightarrow \l'=-{1\over \l}$ and get the formula
\beq\label{nnf}
E_0^c(\l')= -{1\over 4-C}+\sum_{k=1}{\rho_k'(C)\over {\l_k}'(C)-{\l}'}. 
\eeq

After that  the  proof of the theorem  follows familiar steps. The Poisson bracket for the function $E_0^c(\l')$ is given by  formula \ref{ahb}.
For the variables $\rho'(C)$ and $\l'(C)$ from \ref{nnf}, an  analog of Lemma \ref{ocp} holds.  Then  similarly  to 
Lemma \ref{oocb} we can compute the bracket between  the variables $\rho(C)$ and $\l(C)$ entering 
into  formula \ref{nf}. Finally, the proof of the canonical relations is identical to the proof of Theorem \ref{tcp}.  
$
\blacksquare
$

We will construct a family of Hamiltonian flows which will be linearized in the variables $I(C)$ and $\theta(C)$ in section 11.

\section{Second family of canonical  coordinates}

The arguments  of the previous section prompt  the construction of  another family of canonical coordinates  $J$ and $\tau$.  As it was explained in the introduction   these  variables are complimentary to $I$ and $\t$. We assume that the initial data belong to the smaller phase 
space $\MM_0$.

According to formula \ref{gggg},  the function $E_0(\l)$ has the form
$$
E_0(\l)=-{\varphi(2,\l)\over \psi(2,\l)}=-{1\over l} +\sum_{k=1}^{\infty} {\rho_k\over \l_k-\l}.
$$
The range of all possible sequences of $\l$'s is described in Theorem \ref{krlia}.

Pick some $F$  such that  $-{1\over l} <F$ and consider the points where $E_0(\l)=F$ 
(see Figure 2).
We have the sequence of points $\mu_k=\mu_k(F),\;k=1,2,\hdots.$ interlacing the poles of 
$E_0(\l)$:
\beq\label{inm}
\mu_1<\l_1<\mu_2<\l_2<\hdots.
\eeq
Given $\l$'s, how can the range of all possible sequences of $\mu(F)$'s be described?

 The Nazarov--Yuditskii theorem for this case implies 
$$
\sum_{k=1}^{\infty} \(\frac{\l_k}{\mu_k(F)}-1\) <\infty.
$$
Conversely, suppose the sequence of $\mu$'s satisfies \ref{inm} and the above condition. Then again, by the Nazarov--Yuditskii theorem, there exists a 
function 
$$
\U(\l)=\sum_{k=1}^{\infty} \frac{\rho_k^0}{\l_k-\l},\qquad\qquad\qquad\sum_{k=1}^{\infty} \frac{\rho_k^0}{\l_k} <\infty;
$$
such that $\U(\mu_k)=1,\quad k=1,2,\hdots$. Note that $\l_1 >0$ and $\U(0)>0$  depends only on the chosen sequence of $\mu$'s and $\l$'s. Pick $l,\; 
0< l< 4$ and define 
\beq\label{ipps}
E_0(\l)=-\frac{1}{l}+ \frac{(4-l)\U(\l)}{4l\,\U(0)}.
\eeq
By construction
$$
E_0(\mu_k)=-\frac{1}{l}+ \frac{(4-l)}{4l\,\U(0)} =F,\qquad\qquad\qquad k=1,2,\hdots.
$$
For the range of $F$ we have three  options. 
\begin{itemize} 
\item If $\mu_1 >0$ then $\U(0)< 1$ and  F satisfies   $-1/4<  F < +\infty $.
\item If $\mu_1 < 0$ then $\U(0)> 1$ and  F satisfies   $-\infty <  F < -1/4$.
\item If $\mu_1 =0$ then $\U(0)=1$ and  $F=-1/4$.
\end{itemize}
The function $E_0(\l)$ has  the form \ref{gg} and satisfies  condition \ref{ggh}. Therefore, according to remark \ref{weylspec} for any choice of $l$,
 or equivalently $F$,  
there exists a string $\S_0$ with the function $E_0$ given  by  formula \ref{ipps}.

\begin{rem} 
Let  $\mu_1 >0$ and $F=0$. Then there  exists a string $\S_0$ with $l>0$ and corresponding  spectrum $\mu(0)$'s. 
\end{rem}

The next theorem is an  analog of Theorem \ref{ttcp}. We pick $F$ such that $\mu_1(F)\neq 0$.

\begin{thm} The variables 
$$
J_n(F)=-{1\over \mu_n(F)},\qquad\qquad\qquad \tau_k(F)=\log \psi(2,\mu_k(F))
$$
are canonically conjugate variables
\bey
\{J_k(F),J_n(F)\}&=&0,\\
\{\tau_k(F),\tau_n(F)\}&=&0,\\
\{\tau_k(F),J_n(F)\}&=&\delta_k^n.\\
\eey
\end{thm}

To prove  the theorem we introduce the function $\Omega_0^F(\l)$:
$$
\Omega_0^F(\l)=-{1\over E_0(\l)-F}.
$$
It can be written as 
$$
\Omega_0^F(\l)={\psi\over \varphi + F \psi}={\psi\over \varphi^F}.
$$
The points  $\mu_k(F)$ are the roots of the equation $\varphi^F(\l)=0$. 
The function $\varphi^F$ can be written as:
$$
\varphi^F(\l)=(1+4F)\prod_{k=1}^{\infty}\(1-{\l\over \mu_k(F)}\). 
$$
The function $\Omega_0^F(\l)$ maps the upper half--plane into itself and has the  expansion
\beq\label{abs}
\Omega_0^F(\l)={l\over 1+ Fl} +\sum_{k=1}^{\infty}{\sigma_k(F)\over \mu_k(F)-\l}.
\eeq
One can change the variable $\l\rightarrow \l'= -{1\over \l}$  and  get the formula
\beq\label{aabs}
\Omega_0^F(\l')={4\over 1 + 4F} + \sum_{k=1}^{\infty}{\mu'_k(F)\over \sigma'_k(F)-\l'}.
\eeq

After that  the  proof of the theorem  follows familiar steps. The Poisson bracket for the function $\Omega_0^F(\l')$ is given by  formula \ref{ahb}.
For the variables $\sigma'(F)$ and $\mu'(F)$ from \ref{aabs} an  analog of Lemma \ref{ocp} holds.  Then  similarly to 
Lemma \ref{oocb} we can compute the bracket between  the variables  $\sigma(F)$ and $\mu(F)$ entering 
into  formula \ref{abs}. Finally, the proof of the canonical relations is identical to the proof of Theorem \ref{tcp}.  
$
\blacksquare
$

We will construct a family of Hamiltonian flows which will be linearized in the variables $J(F)$ and $\tau(F)$ in the next section.

\section{Trace formulas}
Let  $\l_k(a,b)$ be a  spectrum of the boundary value problem 
$$
a\varphi(2,\l)+b \psi(2,\l)=0.
$$
In this section we will express the sums 
$$
\sum_{k=1}^{+\infty} \frac{1}{\l_k(a,b)}, \qquad\sum_{k=1}^{+\infty} \frac{1}{\l_k^2(a,b)}, \qquad etc;
$$
in terms of the  potential. In order to simplify calculations we work with  the Sturm--Liouville problem \ref{slsp}. 
We  assume that the potential 
$m(x)$ has compact support.  The solutions $\varphi(x,\l)$ and $\psi(x,\l)$   of the Sturm--Liouville problem are the image of the corresponding solutions of the string problem under the Liouville 
transformation. Note 
$$
\varphi(x,0)=\varphi_0(x)=\cosh x/2,\qquad\qquad \psi(x,0)=\psi_0(x)=2e^{x/2}.
$$ 
We also introduce the solution $\theta(x,\l)$ which corresponds to the solution $\theta(\xi,\l)$  specified by the 
boundary condition at the right end
$$
\theta(+2,\l)=0,\qquad\qquad\qquad\qquad \theta'_+(+2,\l)=1.
$$
Note that 
$$
\theta(\xi,0)=\xi-2,\qquad\qquad\qquad \theta(x,0)=\theta_0(x)=2e^{-x/2}.
$$  
At the points of the spectrum $\l_k(a,b)$,  the solution $\Delta(x,\l)=a\varphi(x,\l)+b \psi(x,\l)$ tends to zero  when $x$ tends $+\infty$. At these points, the Wronskian 
$$
W(\l)=\theta(x,\l)D\Delta(x,\l)- \Delta(x,\l) D\theta(x,\l),\qquad\qquad 
$$ 
vanishes.

The expansion can be obtained   near $\l=0$ 
$$
W(\l)=I_0(a,b) +I_1(a,b)\l +I_2(a,b)\l^2+\hdots
$$
in two different ways. The first arises from the Hadamard formula
$$
W(\l)=(a+4b) \prod_{k=1}^{\infty}\(1-\frac{\l}{\l_k}\),\qquad\qquad\qquad \l_k=\l_k(a,b).
$$
The coefficients have  the form
\bey
I_0(a,b) &=& a+4b,\\
I_1(a,b)&=& -(a+4b)\sum_{k}\frac{1}{\l_k},\\
I_2(a,b)&=&  (a+4b)\sum_{n\neq k}\frac{1}{\l_k \l_n}.
\eey
Another expansion is obtained from the Neumann series for the solutions 
\bey
\varphi(x,\l)&=& \varphi_0(x) +\varphi_1(x)\l + \varphi_2(x)\l^2 +\hdots;\\
\psi(x,\l)&=& \psi_0(x) +\psi_1(x)\l + \psi_2(x)\l^2 +\hdots ;\\
\theta(x,\l)&=& \theta_0(x) +\theta_1(x)\l + \theta_2(x)\l^2+\hdots. 
\eey
Therefore,  
\bay
I_0(a,b) &=& \theta_0 D\Delta_0- D\theta_0 \Delta_0,\label{coefone}\\
I_1(a,b)&=& \theta_0 D\Delta_1 + \theta_1 D\Delta_0-D\theta_0 \Delta_1-D\theta_1 \Delta_0 ,\label{coeftwo}\\
I_2(a,b)&=& \theta_0 D\Delta_2 + \theta_1 D\Delta_1+\theta_2 D\Delta_0 -D\theta_0 \Delta_2-D\theta_1 \Delta_1-D\theta_2 \Delta_0;\label{coefthree}
\ey
where $\Delta_k(x)=a \varphi_k(x)+b\psi_k(x)$. 
Identifying the corresponding coefficients we obtain the first two trace formulas 
\bay
\sum_{k=1}^{+\infty} \frac{1}{\l_k(a,b)}&=& -\frac{I_1(a,b)}{a+4b},\label{firsttrf}\\
\sum_{k=1}^{+\infty} \frac{1}{\l_k^2(a,b)}&=& \[\frac{I_1(a,b)}{a+4b}\]^2 -\frac{2I_2(a,b)}{a+4b}.\label{secondtrf}
\ey

The coefficients $\varphi_k(x)$ and $\psi_k(x)$ are given  by the formulas
\bey
\varphi_0(x)&=&\cosh \frac{x}{2},\\
\varphi_1(x)&=&-2\int\limits_{-\infty}^{x}d\xi\, \sinh\frac{x-\xi}{2}\cosh\frac{\xi}{2}\, m(\xi),\\
\varphi_2(x)&=&4\int\limits_{-\infty}^{x}d\xi\int\limits_{-\infty}^{\xi} d\eta \, \sinh\frac{x-\xi}{2}\sinh\frac{\xi-\eta}{2} \cosh\frac{\eta}{2} \,m(\xi)m(\eta),\\
\psi_0(x)&=&2e^{x/2},\\
\psi_1(x)&=&-4\int\limits_{-\infty}^{x}d\xi\, \sinh\frac{x-\xi}{2}\,e^{\xi/2}\, m(\xi),\label{coone}\\
\psi_2(x)&=&8\int\limits_{-\infty}^{x}d\xi\int\limits_{-\infty}^{\xi} d\eta \, \sinh\frac{x-\xi}{2}\sinh\frac{\xi-\eta}{2}\, e^{\eta/2} \,m(\xi)m(\eta).\label{cotwo}
\eey
Also,
\bey
\theta_0&=&2e^{-x/2},\\
\theta_1&=&4\int\limits_{x}^{+\infty} d\xi \sinh\frac{x-\xi}{2}\e^{-\xi/2}m(\xi),\\
\theta_2&=& 8\int\limits_{x}^{+\infty}d\xi\int\limits_{\xi}^{+\infty} d\eta \sinh\frac{x-\xi}{2}\sinh\frac{\xi-\eta}{2}e^{-\eta/2}m(\xi)m(\eta).    
\eey
It is easy to check that formula \ref{coefone} produces $I_0(a,b)=a+4b$. 
Note that the coefficients $I_k(a,b)$ are linear in the parameters $a$ and $b$:
$I_k(a,b)=aI_k(1,0) +bI_k(0,1).$  
For the first trace formula \ref{firsttrf} we have
$$
\sum_{k=1}^{+\infty} \frac{1}{\l_k(a,b)}=-\frac{a I_1(1,0)}{a+4b}- 
\frac{bI_1(0,1)}{a+4b}. 
$$
From  \ref{coeftwo}  with the help of  formulas for the coefficients of expansions of the solutions we have  
\bey
I_1(1,0)&=& -\int_{-\infty}^{+\infty} dx\, m(x)(1+e^{-x}),\\
I_1(0,1)&=& -4 \int_{-\infty}^{+\infty} dx \,m(x). 
\eey

For the second  trace formula \ref{secondtrf} we have
$$
\sum_{k=1}^{+\infty} \frac{1}{\l_k^2(a,b)}=\(\frac{a I_1(1,0)+ bI_1(0,1)}{a+4b}\)^2- 
2\times \frac{ a I_2(1,0)+ bI_2(0,1)}{a+4b}. 
$$
From  \ref{coefthree}  with the help of  the formulas for the coefficients of the expansions  we have
\bey
I_2(1,0)&=&\frac{1}{2}\(\;\;\int\limits_{-\infty}^{+\infty} dx\, m(x)\;\,\)^2-2 \int\limits_{-\infty}^{+\infty} dx\, e^{-x}\,m(x) D^{-1}m(x) -\int\limits_{-\infty}^{+\infty} dx\, m(x) v(x),\\
I_2(0,1)&=&2\(\;\;\int\limits_{-\infty}^{+\infty} dx\, m(x)\;\,\)^2 -4 \int\limits_{-\infty}^{+\infty} dx\, m(x) v(x). 
\eey

The first series of Hamiltonians $\HH_k(C)$ corresponds to the case $a=-C,\; b=1$. Two Hamiltonians $\HH_1(C)$ and $\HH_2(C)$ are given in 
 1.4.   The second series of Hamiltonians $\TT_k(F)$ corresponds to the case $a=1,\; b=F$. The first few also can be found in 1.4.

\

\vskip .2in
\noindent
Department of Mathematics
\newline
Michigan State University
\newline
East Lansing, MI 48824
\newline
USA
\vskip 0.3in
\noindent
vaninsky@math.msu.edu

\end{document}